\documentclass[twocolumn,prl,reprint,twocolumn,superscriptaddress,amsmath,amssymb]{revtex4-1}
\usepackage[latin9]{inputenc}
\usepackage{amsmath}
\usepackage{graphicx}
\usepackage[unicode=true,pdfusetitle,
 bookmarks=false,
 breaklinks=false,pdfborder={0 0 1},backref=false,colorlinks=false]
 {hyperref}

\makeatletter

\DeclareTextSymbolDefault{\textquotedbl}{T1}

\usepackage{color}
\usepackage{times}
\usepackage[caption=false]{subfig}

\usepackage{graphicx}

\makeatother

\begin{document}
\title{Topology by Dissipation: Transport properties }
\author{Gal Shavit}
\affiliation{Raymond and Beverly Sackler School of Physics and Astronomy, Tel Aviv
University, Tel Aviv 6997801, Israel}
\author{Moshe Goldstein}
\affiliation{Raymond and Beverly Sackler School of Physics and Astronomy, Tel Aviv
University, Tel Aviv 6997801, Israel}
\begin{abstract}
Topological phases of matter are the center of much current interest,
with promising potential applications in, e.g., topologically-protected
transport and quantum computing. Traditionally such states are prepared
by tuning the system Hamiltonian while coupling it to a generic bath
at very low temperatures. However, this approach is often ineffective,
especially in cold-atom systems. It was recently shown that topological
phases can emerge much more efficiently even in the absence of a Hamiltonian,
by properly engineering the interaction of the system with its environment,
to directly drive the system into the desired state. Here we concentrate
on dissipatively-induced 2D Chern insulator (lattice quantum Hall)
states. We employ open quantum systems tools to explore their transport
properties, such as persistent currents and the conductance in the
steady state, in the presence of various Hamiltonians. We find that,
in contrast with equilibrium systems, the usual confinement of currents
to the edge, as well as the usual relation between the Chern topological
number and the Hall conductance could be broken. We explore the intriguing
edge behaviors and elucidate under which conditions the Hall conductance
is quantized.
\end{abstract}
\maketitle
\section {Introduction} 

The adverse affects of a dissipative environment or bath on a quantum
system coupled to it are well-known: Processes of decay and decoherence
lead to a degradation of the desired quantum state and shortening
of its coherence time \citep{quantumNoise,quantumComputation}. For
these reasons, a prime goal of quantum device engineering is to mitigate
those effects, mainly by trying to reduce the system-bath interaction
strength. However, it has been realized that an \textit{engineered}
dissipative interaction to an external bath can possibly be exploited,
in order to reliably prepare interesting non-trivial quantum states
in open systems; This approach has recently been brought to many-body
systems \citep{DisPrep1,DisPrep2,Disprep6,DisPrep3,DisPrep4,DisPrep7,DisPrep5}.
This is typically done by tailoring the bath interaction such that
the required quantum state would emerge as a unique steady state solution
to the open quantum (Lindblad \citep{lindbladSemigroups}) master
equation, independent of the initial conditions. If the sought after
state is represented by the density matrix $\rho_{D}$, one would
aim to manipulate the system-bath coupling such that $\rho_{D}$ is
a ``dark state'' of the dynamical evolution of the open system,
i.e.,
\begin{equation}
\frac{d}{dt}\rho_{D}=0\label{eq:darkDensityMatrix}
\end{equation}

This idea of dissipative preparation contrasts the conventional Hamiltonian
approach, that heavily relies on reaching sufficiently low temperatures
to attain the interesting properties of the desired quantum ground
state. This approach might prove ineffective in some cases. Of particular
interest is the case of quantum simulators implemented using ultra-cold
atomic gases \citep{Quantumsimulation,ultracold25}, where equilibrium
at low temperatures compared to the trapping potential energy scale
is experimentally challenging to achieve. A dissipative preparation
scheme, albeit with its own challenges and complexities, may circumvent
such issues, as the accuracy of the final state will be determined
solely by the degree of control one has on the details of the engineered
bath coupling.

Topologically non-trivial phases of matter \citep{TopoOrderWen} have
been at the forefront of condensed matter physics during the last
several decades, since the discovery of the integer and fractional
quantum Hall effects \citep{IntroFQHexp1,IntroFQHexp2}. The more
recently discovered topological insulators \citep{TOPO1,TOPO2,TOPO3},
superconductors \citep{TOPO4,Lutchyn2018}, and semimetals \citep{Weyl}
have opened the door to novel exciting possibilities, e.g., topologically
protected quantum computing \citep{QuantComp}. The idea of employing
the engineered bath interactions scheme to stabilize a topologically
ordered ground state has already received some attention in recent
years \citep{disstopom1,disstopo0p5,MG8,MG9,MG11,DisPrep5,disstopo0,MG13,MG14,MG}.

We focus on the 2D Chern insulator phase discussed in Ref. \citep{MG},
stabilized by employing \textit{purely dissipative} dynamics of Lindblad
type, i.e., the master equation describing the evolution of the system
density matrix $\rho$,
\begin{equation}
\frac{d}{dt}\rho=-i\left[H,\rho\right]+{\cal D}\rho,\label{eq:Lindbladian}
\end{equation}
has the Hamiltonian $H=0$, and where ${\cal D}$ is some general
engineered Lindbladian dissipator super-operator acting on $\rho$.
${\cal D}$ is of the form $\mathcal{D}\rho=\sum_{j}\gamma_{j}\left(L_{j}\rho L_{j}^{\dagger}-\frac{1}{2}\left\{ \rho,L_{j}^{\dagger}L_{j}\right\} \right)$,
with $L_{j}$'s being a set of ``quantum jump'' operators, and $\gamma_{j}$'s
the rates governing the dissipative dynamics \citep{quantumNoise}.
The jump operators and rates which we use in this work will be specified
later on. This work addresses the effects of including $H\neq0$ in
the Lindblad dynamics \eqref{eq:Lindbladian}. This step is crucial,
as it allows one to define current operators in the system, such that
transport properties of the dissipative state may be explored. These
properties, e.g, the Hall conductance, are the hallmark of the well-known
equilibrium counterpart of this phase, and as such are important in
characterizing the engineered state. The interplay of Hamiltonian
and Lindbladian dynamics in topological systems has also been the
focus of some recent works \citep{AdditionalReference1,AdditionalReference2}.
We will show that although the dissipatively engineered states could
be arbitrarily close to equilibrium quantum Hall states, they do not
always present the same persistent current and transport features.
These depend on the details of the coherent dynamics, the relative
amplitude of those dynamics compared to the dissipation energy scale,
and on the implementation of artificial fields within the dissipative
scheme.

The rest of the paper is organized as follows. In Sec. \ref{dissScheme}
we briefly introduce the dissipative scheme developed in \citep{MG}.
We introduce the Hamiltonian dynamics, and the tools required to analyze
the steady state in its presence in Sec. \ref{Hintroduction}. This
step enables us to discuss the persistent steady state currents that
develop for different classes of Hamiltonians in Sec. \ref{perestJ}.
Then, the electric conductance response is calculated in Sec. \ref{conductivity}
for the different cases, and compared with the known results for the
equilibrium scenario. Finally, we summarize our findings and conclusions
in Sec. \ref{conclusions}. Some technical details are relegated to
the Appendices.

\begin{figure}
\begin{centering}
\includegraphics[bb=48.0418bp 0bp 552bp 457bp,scale=0.45]{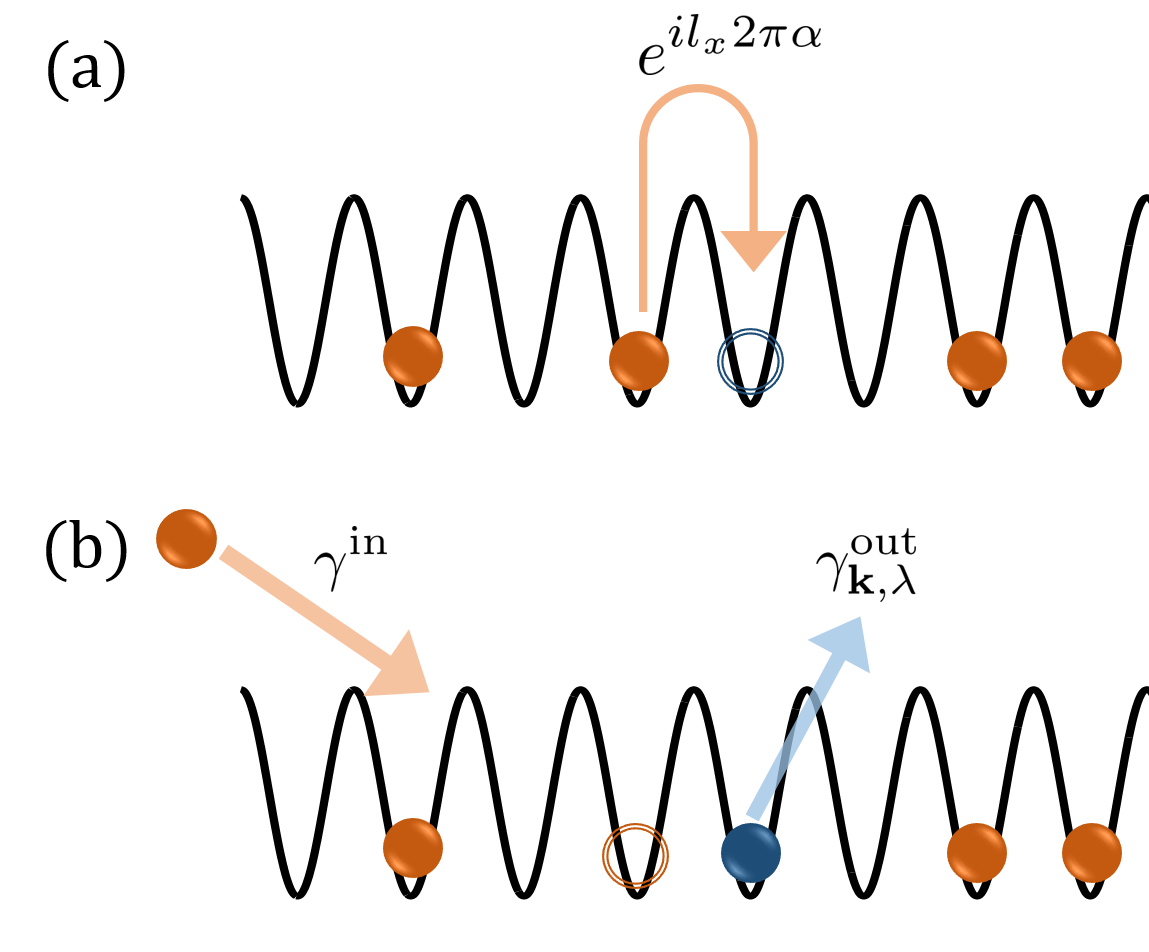}
\par\end{centering}
\caption{\label{fig:Schematics} Schematic depiction of the dissipative scheme
described in the text. (a) Orange particles, representing the $a$
particles, may hop to a nearest neighbor site and become a $b$ particle
(blue). This hopping has a phase which is determined by the artificial
magnetic field implemented in the lattice, according to Eq. \eqref{eq:dissH}.
(b) The $b$ particles are not trapped, and escape the lattice, which
imposes a band and momentum dependent effective depletion rate for
the $a$ particles, $\gamma_{{\bf k},\lambda}^{{\rm out}}$. An external
refilling reservoir for $a$ is added, with a tuned wavevector- and
band-independent refilling rate $\gamma^{{\rm in}}$.}
\end{figure}

\section {Dissipatively induced topological state}\label{dissScheme}

\begin{figure}
\begin{centering}
\includegraphics[bb=45bp 0bp 879bp 405bp,scale=0.3]{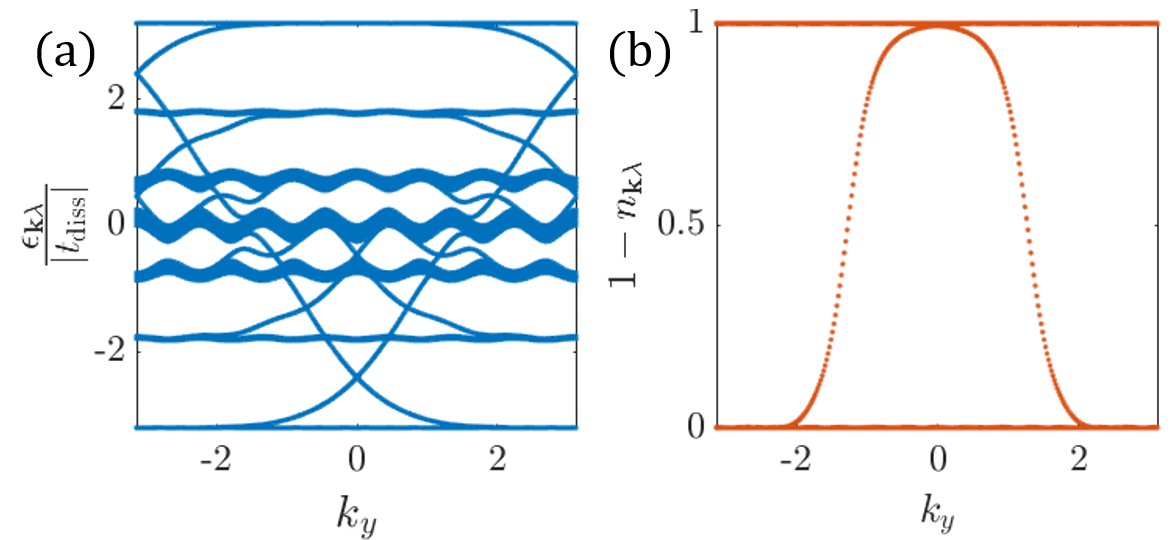}
\par\end{centering}
\caption{\label{fig:FigDissQH} Example of the purely-dissipative scheme. (a)
Calculated spectrum $\epsilon_{\mathbf{k},\lambda}$ of the reference
Hamiltonian. The different bands are visibly shown, with the edge
state spectrum appearing in the interband gaps. (b) Using the reference
Hamiltonian to construct $H^{{\rm diss}}$, the steady state occupation
numbers are obtained. The bottom band is almost entirely populated,
whereas all the rest are nearly depleted (The deviation from population
values of 0 and 1 is $\sim0.002$). We set $\mu^{*}=\overline{\epsilon_{\mathbf{k},1}}\equiv\frac{1}{L_{x}L_{y}}\sum_{{\bf k}}\epsilon_{\mathbf{k},1}$,
and $\gamma_{\mathrm{opt}}^{\mathrm{in}}$ was calculated according
to Eq. \eqref{eq:Kopr}. In both plots we used $\alpha=\frac{1}{7}$,
$L_{x}=L_{y}=301$, periodic boundary conditions along the $y$ direction,
and open boundary conditions in the $x$ direction (cylindrical geometry).}
\end{figure}

In this section, we recapitulate the dissipative recipe presented
in Ref. \citep{MG} to realize a dissipative lattice integer quantum
Hall state. The main components in the scheme are illustrated in Fig.
\ref{fig:Schematics}. Consider the Harper-Hofstadter model \citep{HarperModel1955,Hofstadter1976},
describing nearest-neighbor hopping on a two-dimensional square lattice
pierced by a magnetic flux, which we will use as a reference in the
construction below, 
\begin{align}
H^{{\rm ref}} & =-t_{{\rm ref}}\sum_{l_{x},l_{y}}a_{l_{x},l_{y}}^{\dagger}\left(e^{il_{x}2\pi\alpha}a_{l_{x},l_{y}+1}+a_{l_{x}+1,l_{y}}\right)+\mathrm{h.c.\,,}\label{eq:refH}
\end{align}
with $a_{l_{x},l_{y}}$ a fermionic annihilation operator on the site
$\left(l_{x},l_{y}\right)$, $t_{{\rm ref}}$ the hopping strength,
and $\alpha$ is the magnetic flux per plaquette in units of flux
quantum, introduced using the Peierls substitution. $\alpha$ is assumed
to be a rational fraction $\alpha=\frac{p}{q}$ (with $p,q,$ integers
with no common factor). The Hamiltonian \eqref{eq:refH} is diagonalized
by moving to two-dimensional momentum space,
\begin{equation}
H^{{\rm ref}}=\sum_{\mathbf{k},\lambda}\epsilon_{\mathbf{k},\lambda}a_{\mathbf{k},\lambda}^{\dagger}a_{\mathbf{k},\lambda},\label{eq:refHklambda}
\end{equation}
with ${\bf k}$ the two-dimensional momentum, and $\lambda=1,2,...,q$
the band index. The spectrum obtained for $\alpha=\frac{1}{7}$ with
periodic boundary conditions along one direction is shown in Fig.
\ref{fig:FigDissQH}(a), where the distinct bands are apparent, as
well as the edge state spectrum.

The goal of the construction of \citep{MG} is to achieve a state
as close as possible to the ground state of the reference Hamiltonian
with chemical potential in the first gap (i.e., the lowest band completely
filled and all the other bands completely empty) by purely dissipative
dynamics. For that we do not implement the Hamiltonian \eqref{eq:refH}--\eqref{eq:refHklambda},
but rather introduce a different Hamiltonian, built using the matrix
elements of the Hofstadter reference Hamiltonian \eqref{eq:refH},

\begin{align}
H^{{\rm diss}} & =-t_{{\rm diss}}\sum_{l_{x},l_{y},s=\pm1}b_{l_{x},l_{y}}^{\dagger}\left(e^{isl_{x}2\pi\alpha}a_{l_{x},l_{y}+s}+a_{l_{x}+s,l_{y}}\right)\nonumber \\
 & -\mu^{*}\sum_{l_{x},l_{y}}b_{l_{x},l_{y}}^{\dagger}a_{l_{x},l_{y}}+\mathrm{h.c.}\,,\label{eq:dissH}
\end{align}
describing a coupling of the lattice fermions $a$ to an additional
fermionic species $b$ (e.g., a different hyperfine state). A thorough
discussion regarding the cold-atom implementation for this Hamiltonian
is presented in \citep{MG}, and is based on previous equilibrium
constructions for the realization of artificial gauge fields \citep{ZollerImplementation,GerbierColdAtomsGauge}.
In the eigenbasis of the reference Hamiltonian we obtain 
\begin{equation}
H^{{\rm diss}}=\sum_{\mathbf{k},\lambda}\left(\epsilon_{\mathbf{k},\lambda}-\mu^{*}\right)b_{\mathbf{k},\lambda}^{\dagger}a_{\mathbf{k},\lambda}+{\rm h.c.}\,,\label{eq:Hdissjlambda}
\end{equation}
with $\epsilon_{\mathbf{k},\lambda}$ now proportional to $t_{{\rm diss}}$
instead of $t_{{\rm ref}}$. Note that now they do not represent energies
in the $\mathbf{k},\lambda$ single particle basis (since we do not
implement reference Hamiltonian \eqref{eq:refH}--\eqref{eq:refHklambda},
though we still use its eigenbasis to define the state). Rather, they
represent the amplitudes of the coupling of the $a$ and $b$ particle
in the $\mathbf{k},\lambda$ state. 

We assume that the $a$-particles are optically trapped in the 2D
system plane, whereas the $b$-particles are free to escape along
the perpendicular $z$ direction. This dynamics of the $b$-particles
is captured by supplementing $H^{{\rm diss}}$ by an Hamiltonian for
the bath particles, $H^{{\rm bath}}=\sum_{l_{x},l_{y},q_{z}}\epsilon_{z}\left(q_{z}\right)b_{l_{x},l_{y},q_{z}}^{\dagger}b_{l_{x},l_{y},q_{z}}$,
where $q_{z}$ is the momentum of the $b$-particles along the $z$
direction, and $\epsilon_{z}\left(q_{z}\right)$ the corresponding
dispersion relation. The $a$-particles are coupled via Eq. \eqref{eq:dissH}
to the $b$ operators in the system plane $z=0$, i.e., to $b_{l_{x},l_{y}}=\sum_{q_{z}}\frac{1}{\sqrt{N_{z}}}b_{l_{x},l_{y},q_{z}}$,
where $N_{z}$ is the system size in the perpendicular direction.
We assume that, in the cold-atom implementation, $b$-particles which
leave the 2D lattice escape to infinity, and as a consequence, $\left\langle b_{l_{x},l_{y},q_{z}}^{\dagger}b_{n_{x},n_{y},q_{z}^{\prime}}\right\rangle ={\rm Tr}\left\{ \rho_{B}b_{l_{x},l_{y},q_{z}}^{\dagger}b_{n_{x},n_{y},q_{z}^{\prime}}\right\} =0$,
for all $q_{z}$, $q_{z}^{\prime}$, with $\rho_{B}$ the density
matrix of the bath particles. Hence, the $b$-particles constitute
a bath with infinitely negative chemical potential for the $a$-particles,
and we can integrate them out. 

Here the central point of this construction becomes clear: The Hamiltonian
of the $a$ particles themselves is zero; they only experience a coupling
to the $b$-particle bath. Hence, the dynamics is not set by the energetics
of the $a$ system, but rather via controlling the matrix elements
of the $a$-$b$ coupling by Eqs. \eqref{eq:dissH}--\eqref{eq:Hdissjlambda}.
As detailed in Ref. \citep{MG}, upon integrating out the $b$ bath,
one arrives at a contribution to the Lindblad master equation for
the density matrix of the $a$ particles, 
\begin{align}
\mathcal{D}^{\mathrm{out}}\rho & =\sum_{\mathbf{k},\lambda}\gamma_{\mathbf{k},\lambda}^{\mathrm{out}}\left(a_{\mathbf{k},\lambda}\rho a_{\mathbf{k},\lambda}^{\dagger}-\frac{1}{2}\left\{ a_{\mathbf{k},\lambda}^{\dagger}a_{\mathbf{k},\lambda},\rho\right\} \right).
\end{align}
Calculation of the rates $\gamma_{\mathbf{k},\lambda}^{\mathrm{out}}$
can be done using Fermi's golden rule \citep{DiracGoldenRule}, leading
to
\begin{equation}
\gamma_{\mathbf{k},\lambda}^{\mathrm{out}}=\frac{2\pi}{\hbar}\nu_{0}\left|\epsilon_{\mathbf{k},\lambda}-\mu^{*}\right|^{2},\label{eq:gamma_outs}
\end{equation}

\noindent with $\nu_{0}$, the density of available $b$-states, taken
as constant \citep{MG}. We thus define a typical dissipative scale
for the system,
\begin{equation}
\gamma^{0}\equiv\frac{2\pi}{\hbar}\nu_{0}\left|t_{\mathrm{diss}}\right|^{2}.\label{eq:gamma0}
\end{equation}
 The implication of \eqref{eq:gamma_outs} is that given a flat band,
i.e., $\epsilon_{\mathbf{k},\lambda}=\epsilon_{0}$ for one value
of $\lambda$, it is possible to fine-tune $\mu^{*}$ such that the
depletion rate goes to zero for that particular band alone. The system
will exponentially decay into a state where the finely-tuned flat
band is the only occupied band, with its occupancy depending on the
initial state of the lattice filling. This is analogous to a low temperature
equilibrium scenario where the chemical potential lies in a gap between
the bands. Again, this is achieved by tuning the system-bath coupling
matrix elements, not their energetics.

Since a topologically non-trivial (non-zero Chern number) exactly
flat-band in a finite-range hopping Hamiltonian, separated by a finite
gap from the other bands, is not possible \citep{NoFlatChen,NoFlat},
we consider a band which is nearly flat, i.e., with a width much smaller
compared to its minimal distance to the other bands. Then, a very
small value of $\gamma_{\mathbf{k},\lambda}^{\mathrm{out}}$ for that
particular band is attainable. We now introduce another ingredient
to the dissipative scheme, a global filling reservoir, replenishing
all bands at a rate $\gamma^{\mathrm{in}}$. This can be achieved
by coupling the system to another reservoir of $c$-particles with
infinitely positive chemical potential. For simplicity the corresponding
reference Hamiltonian will be taken as proportional to the unit matrix
in either real or $\mathrm{\boldsymbol{k}},\lambda$ space, leading
to $\mathrm{\boldsymbol{k}},\lambda$-independent refilling rate \citep{MG}.
The dissipative master equation now reads
\begin{align}
{\cal D}\rho & =\sum_{\mathbf{k},\lambda}\gamma_{\mathbf{k},\lambda}^{\mathrm{out}}\left(a_{\mathbf{k},\lambda}\rho a_{\mathbf{k},\lambda}^{\dagger}-\frac{1}{2}\left\{ a_{\mathbf{k},\lambda}^{\dagger}a_{\mathbf{k},\lambda},\rho\right\} \right)\nonumber \\
 & +\gamma^{\mathrm{in}}\sum_{\mathbf{k},\lambda}\left(a_{\mathbf{k},\lambda}^{\dagger}\rho a_{\mathbf{k},\lambda}-\frac{1}{2}\left\{ a_{\mathbf{k},\lambda}a_{\mathbf{k},\lambda}^{\dagger},\rho\right\} \right).\label{eq:fullMaster}
\end{align}
The steady-state band occupation numbers, $n_{{\bf k},\lambda}={\rm Tr}\left\{ \rho a_{\mathbf{k},\lambda}^{\dagger}a_{\mathbf{k},\lambda}\right\} $
can now be calculated,
\begin{equation}
n_{{\bf k},\lambda}=\frac{\gamma^{\mathrm{in}}}{\gamma^{\mathrm{in}}+\gamma_{\mathbf{k},\lambda}^{{\rm out}}}.\label{eq:occDiss}
\end{equation}
Given the values of the maximal rate for the bottom band, $\mathrm{max}\left\{ \gamma_{\mathbf{k},1}^{\mathrm{out}}\right\} $,
and the minimal one for the next (closest) band $\mathrm{min}\left\{ \gamma_{\mathbf{k},2}^{\mathrm{out}}\right\} $,
an optimal choice of the tunable refilling rate, such that $n_{{\bf k},1}\sim1$
and $n_{{\bf k},\lambda>1}\sim0$ is thus
\begin{equation}
\gamma_{\mathrm{opt}}^{\mathrm{in}}=\sqrt{\mathrm{max}\left\{ \gamma_{\mathbf{k},1}^{\mathrm{out}}\right\} \cdot\mathrm{min}\left\{ \gamma_{\mathbf{k},2}^{\mathrm{out}}\right\} }.\label{eq:Kopr}
\end{equation}
Fig. \ref{fig:FigDissQH}(b) shows an example of results for this
scheme, where we have properly tuned $\mu^{*}$ and $\gamma^{\mathrm{in}}$,
and achieved almost completely full or empty occupation of the lower
band or upper bands, respectively. This mixed state is not only very
close to the pure ground state of the reference Hamiltonian, but also
shares with it the value -1 of the topological index, the Chern number
\citep{MG}.

\section {Introducing Hamiltonian dynamics}\label{Hintroduction}

We now consider adding some coherent dynamics to the $a$ particles
which live on the lattice. This will modify the master equation \eqref{eq:Lindbladian},
since if $H\neq0$, one must include the commutator term between the
Hamiltonian and the density matrix. One may naturally combine the
dissipative artificial gauge field engineering approach of Ref. \citep{MG}
with the non-dissipative artificial gauge field engineering approach
of Refs. \citep{ZollerImplementation,GerbierColdAtomsGauge}, to independently
set the artificial gauge fields in both parts of Eq. \eqref{eq:Lindbladian}.
Of the many possibilities, in this work, we explore the consequences
of two kinds of hopping Hamiltonians,
\begin{align}
H_{{\rm comp}}^{{\rm sys}} & =-t_{{\rm comp}}\sum_{l_{x},l_{y}}a_{l_{x},l_{y}}^{\dagger}\left(e^{il_{x}2\pi\alpha}a_{l_{x},l_{y}+1}+a_{l_{x}+1,l_{y}}\right)+\mathrm{h.c.}\,,\label{eq:Hcomp}
\end{align}
\begin{align}
H_{{\rm inc}}^{{\rm sys}} & =-t_{{\rm inc}}\sum_{l_{x},l_{y}}a_{l_{x},l_{y}}^{\dagger}\left(a_{l_{x},l_{y}+1}+a_{l_{x}+1,l_{y}}\right)+\mathrm{h.c.}\,,\label{eq:Hinc}
\end{align}
where $H_{{\rm comp}}^{{\rm sys}}$ is compatible with the dissipative
interaction Hamiltonian, i.e., it has the dynamics of particles hopping
on a square lattice with the same flux $\alpha$ as before, and $H_{{\rm inc}}$
is a simple nearest-neighbor hopping Hamiltonian which is incompatible.
Understanding the different consequences for particles with a Hamiltonian,
which is either compatible or incompatible with engineered dissipation,
will enable us to determine how system observables change and the
role of topology in the presence of both dissipative and coherent
(Hamiltonian) dynamics. 

The total Hamiltonian, Eq. \eqref{eq:dissH} combined with either
Eq. \eqref{eq:Hcomp} or \eqref{eq:Hinc} {[}and hence the resulting
Lindblad dynamics \eqref{eq:Lindbladian}{]} are invariant under a
general gauge transformation $U=\exp\left[i\sum_{l_{x},l_{y}}f_{l_{x},l_{y}}(a_{l_{x},l_{y}}^{\dagger}a_{l_{x},l_{y}}+b_{l_{x},l_{y}}^{\dagger}b_{l_{x},l_{y}})\right]$.
For example, choosing $f_{l_{x}l_{y}}=-\pi\alpha l_{x}l_{y}$ to go
from the Landau to the symmetric gauge in Eq. \eqref{eq:dissH} leads
to \begin {widetext}
\begin{equation}
H_{S}^{{\rm diss}}=U^{\dagger}H_{L}^{{\rm diss}}U=-t_{{\rm diss}}\sum_{l_{x},l_{y},s=\pm1}b_{l_{x},l_{y}}^{\dagger}\left(e^{isl_{x}\pi\alpha}a_{l_{x},l_{y}+s}+e^{-isl_{x}\pi\alpha}a_{l_{x}+s,l_{y}}\right)-\mu^{*}\sum_{l_{x},l_{y}}b_{l_{x},l_{y}}^{\dagger}a_{l_{x},l_{y}}+\mathrm{h.c.}\,,\label{eq:HdissS}
\end{equation}
The compatible system Hamiltonian then transforms to the symmetric
form as well:
\begin{equation}
H_{{\rm comp},S}^{{\rm sys}}=U^{\dagger}H_{{\rm comp},L}^{{\rm diss}}U=-t_{{\rm comp}}\sum_{l_{x},l_{y}}a_{l_{x},l_{y}}^{\dagger}\left(e^{il_{x}\pi\alpha}a_{l_{x},l_{y}+1}+e^{-il_{y}\pi\alpha}a_{l_{x}+1,l_{y}}\right)+\mathrm{h.c.}\,,\label{eq:HcompS}
\end{equation}
while the incompatible Hamiltonian becomes
\begin{equation}
\tilde{H}_{{\rm inc}}^{{\rm sys}}=U^{\dagger}H_{{\rm inc}}^{{\rm diss}}U=-t_{{\rm inc}}\sum_{l_{x},l_{y}}a_{l_{x},l_{y}}^{\dagger}\left(e^{-il_{x}\pi\alpha}a_{l_{x},l_{y}+1}+e^{-il_{y}\pi\alpha}a_{l_{x}+1,l_{y}}\right)+\mathrm{h.c.}\label{eq:HincTransformed}
\end{equation}
\end {widetext}Of course, gauge invariance dictates that all physical
observables calculated using Eq. \eqref{eq:dissH} and \eqref{eq:Hcomp}
are the same as those calculated with \eqref{eq:HdissS} and \eqref{eq:HcompS}
in the compatible case, and similarly for \eqref{eq:dissH} and \eqref{eq:Hinc}
vs. \eqref{eq:HdissS} and \eqref{eq:HincTransformed} in the incompatible
case. However, another possibility in the latter incompatible case
is to work with Eq. \eqref{eq:HdissS} together with the unmodified
incompatible Hamiltonian \eqref{eq:Hinc}. This latter possibility
(to be referred to as \textquotedbl incompatible-S'' in the following)
is not related by a gauge transformation, and hence physically different
from working with \eqref{eq:dissH} and \eqref{eq:Hinc}, or equivalently
\eqref{eq:HdissS} and \eqref{eq:HincTransformed} (to be referred
to as ``incompatible-L'' in the following). In the following we will
mostly work with the incompatible-L scheme, but also show the difference
with the incompatible-S scheme when relevant. We would like to stress
that both these schemes are equally natural from the point of view
of cold atom implementation.

First, we want to understand how the presence of either \eqref{eq:Hcomp}
or \eqref{eq:Hinc} affects the steady state we arrived at using the
engineered dissipative scheme. We notice right away that the steady
state solution for $\rho$ with $H=0$ commutes with $H_{{\rm comp}}^{{\rm sys}}$,
since it is diagonal in the same ${\bf k},\lambda$ basis as $\gamma^{{\rm out}}$.
As a result, Eq. \eqref{eq:occDiss} holds exactly even for a non-zero
compatible Hamiltonian $H=H_{{\rm comp}}^{{\rm sys}}$.

\begin{figure}
\begin{centering}
\includegraphics[bb=38.3853bp 0bp 523bp 334bp,scale=0.55]{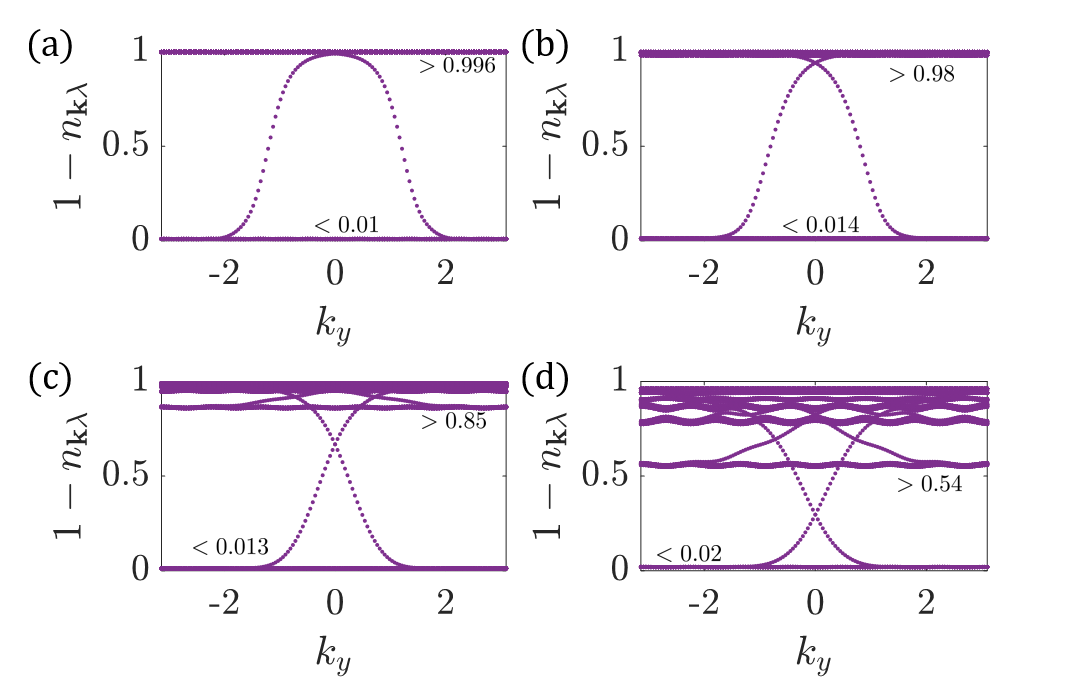}
\par\end{centering}
\caption{\label{fig:tparEffects} Calculation of occupation numbers with $H=H_{{\rm inc}}^{{\rm sys}}$,
see Eq. \eqref{eq:Hinc}. Different values of $t_{{\rm inc}}$ and
$\gamma^{{\rm in}}$ were used in each plot. (a) $t_{{\rm inc}}=0.01\gamma^{0}$,
$\gamma^{{\rm in}}=1.5\gamma_{{\rm opt}}^{{\rm in}}$, (b) $t_{{\rm inc}}=0.03\gamma^{0}$,
$\gamma^{{\rm in}}=10\gamma_{{\rm opt}}^{{\rm in}}$, (c) $t_{{\rm inc}}=0.1\gamma^{0}$,
$\gamma^{{\rm in}}=80\gamma_{{\rm opt}}^{{\rm in}}$, and (d) $t_{{\rm inc}}=0.5\gamma^{0}$,
$\gamma^{{\rm in}}=400\gamma_{{\rm opt}}^{{\rm in}}$. For all plots
we used $\alpha=\frac{1}{7}$, $L_{x}=L_{y}=140$, and the same cylindrical
geometry as in Fig. \ref{fig:FigDissQH}. The largest deviations from
ideal population values of 0 and 1 are indicated within each plot.}
\end{figure}

Conversely, if $H=H_{{\rm inc}}^{{\rm sys}}$ the steady state can
be much different than \eqref{eq:occDiss}, depending on the ratio
$\frac{t_{{\rm inc}}}{\gamma^{0}}$. We use the full master equation
to calculate the steady state single particle density matrix, whose
eigenvalues are the occupation numbers $n_{\mathrm{{\bf k}},\lambda}$,
\begin{equation}
P_{l_{x},l_{y};l_{x}^{\prime},l_{y}^{\prime}}=\mathrm{Tr}\left\{ \rho\left(t\rightarrow\infty\right)a_{l_{x},l_{y}}^{\dagger}a_{l_{x}^{\prime},l_{y}^{\prime}}\right\} .\label{eq:Pdefinitions}
\end{equation}
This requires us to solve a Sylvester-type matrix equation (see Appendix
\ref{SSoccP} for details)
\begin{equation}
\left[\frac{1}{2}\left(\Gamma^{\mathrm{in}}+\Gamma^{\mathrm{out}}\right)+ih\right]P+P\left(\frac{1}{2}\left(\Gamma^{\mathrm{in}}+\Gamma^{\mathrm{out}}\right)-ih\right)=\Gamma^{\mathrm{in}},\label{eq:Psylvester-1}
\end{equation}
with $h$, $\Gamma^{\mathrm{in}}$, and $\Gamma^{\mathrm{out}}$ the
single-particle matrices corresponding to $H$, $\gamma^{\mathrm{in}}$,
and $\gamma^{\mathrm{out}}$, respectively. Eq. \eqref{eq:Psylvester-1}
is similar in essence to the Lyapunov equations derived for the covariance
matrix in Refs. \citep{Sylvester1,Sylvester2,Sylvester3,Lindbladians}.
We find that the engineered nearly-pure steady state deteriorates
with increasing $t_{{\rm inc}}$. As shown in Fig. \ref{fig:tparEffects},
the incompatibility of the Hamiltonian requires one to use a faster
refilling rate $\gamma^{{\rm in}}$ to maintain the high occupation
of the bottom band. This in turn leads to non-negligible occupation
of the higher energy levels, deviating from the desired state. This
phenomenon becomes significant at about $t_{{\rm inc}}\sim0.1\gamma^{0}$,
i.e., when the Hamiltonian is no longer negligible compared to the
dissipative energy scale. We note that the gap in the spectrum of
$P$ (purity gap) is still finite, and the associated Chern number
retains its value of -1 \citep{MG}.

\section {Persistent currents}\label{perestJ}

Including an Hamiltonian for the $a$ species finally allows us to
define a sensible current operator in the system, by looking at the
time evolution of the local particle density expectation value $n_{l_{x},l_{y}}\left(t\right)\equiv{\rm Tr}\left\{ \rho\left(t\right)a_{l_{x},l_{y}}^{\dagger}a_{l_{x},l_{y}}\right\} $
(see Appendix \ref{currentApp} for the full derivation). Using the
master equation \eqref{eq:Lindbladian} for the evolution of the density
matrix, one can separate the local change in particle number into
a coherent contribution $\dot{n}^{H}$ and a dissipative part $\dot{n}^{{\cal D}}$,
\begin{align}
\frac{d}{dt}n_{l_{x},l_{y}}\left(t\right) & =\dot{n}_{l_{x},l_{y}}^{H}\left(t\right)+\dot{n}_{l_{x},l_{y}}^{\mathcal{D}}\left(t\right).\label{eq:dndt20}
\end{align}
The coherent part, controlled by the $\left[H,\rho\right]$ term in
the master equation, must obey a continuity equation involving the
particle current, $\dot{n}^{H}+\nabla\cdot{\bf j}=0$, where $\left(\nabla\cdot\right)$
is a lattice version of the divergence operator, or more explicitly,
\begin{equation}
\dot{n}_{l_{x},l_{y}}^{H}\left(t\right)=-\left(j_{l_{x}+1,l_{y}}^{x}-j_{l_{x},l_{y}}^{x}\right)-\left(j_{l_{x},l_{y}+1}^{y}-j_{l_{x},l_{y}}^{y}\right).
\end{equation}
By examining the expression for $\dot{n}^{H}$, which depends on the
details of the Hamiltonian, one can extract and define a proper current
operator. The current operator which we investigate indeed describes
the local $a$-particle flow rate in the system, and is thus a physical
measurable quantity. As shown in Appendix \ref{currentApp}, the steady
state expectation values of the current  are fully given in terms
of elements of $P$, so no further complicated calculations are required
in order to obtain them. Since in the steady state $\frac{d}{dt}n=0$,
the dissipative steady state contribution can be calculated from the
divergence of the current, $\dot{n}^{{\cal D}}=-\dot{n}^{H}=\nabla\cdot{\bf j}$,
and can be decomposed into incoming and outgoing terms,
\begin{equation}
\dot{n}_{l_{x},l_{y}}^{{\cal D}}\equiv J_{l_{x},l_{y}}^{\mathcal{D},\mathrm{in}}-J_{l_{x},l_{y}}^{\mathcal{D},\mathrm{out}},\label{eq:dndt22}
\end{equation}
which are proportional to the single particle matrix elements of $\Gamma^{{\rm in}}$
and $\Gamma^{{\rm out}}$, respectively (see Appendix \ref{currentApp}).
An alternative derivation for these current operators, which relies
on coupling the Hamiltonian to a probe gauge field, is given at the
end of Appendix \ref{currentApp}. We present our results below for
the different classes of Hamiltonian we have considered.

\subsection {Compatible hopping}

\begin{figure}
\begin{centering}
\includegraphics[bb=18.74101bp 0bp 526.247bp 406.258bp,scale=0.48]{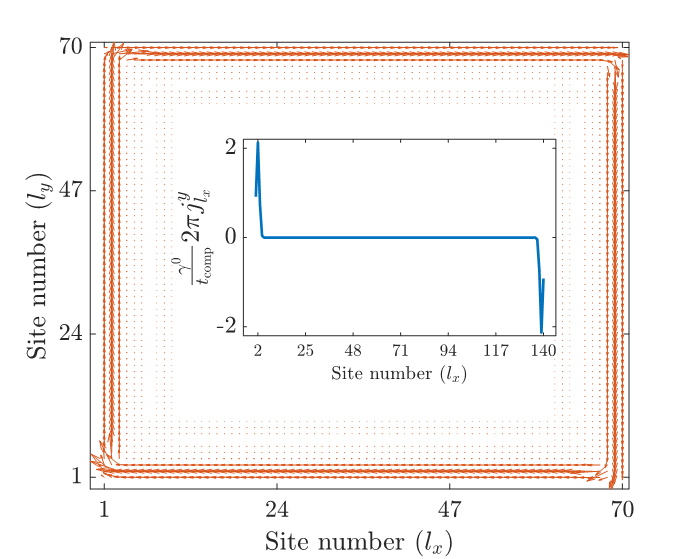}
\par\end{centering}
\caption{\label{fig:compPersist} The current distribution calculated in the
compatible case, Eqs. \ref{eq:dissH} and \ref{eq:Hcomp}, with a
square geometry. Each arrow is proportional to the current density
vector at the corresponding lattice site. We used $\alpha=\frac{1}{7}$,
$L_{x}=L_{y}=70$, and the optimal values for $\mu^{*}$ and $\gamma^{{\rm in}}$.
Localization of a chiral edge mode is visible. Inset: calculation
with a cylindrical geometry (periodic boundaries along the y axis)
of $j_{l_{x}}^{y}=\sum_{l_{y}}j_{l_{x},l_{y}}^{y}$, using the same
parameters, except for $L_{x}=L_{y}=140$.}
\end{figure}

The expectation value of the electric current operator in the different
directions is given in terms of elements of the $P$ matrix by (Appendix
\ref{currentApp})\begin {subequations}\label{jcomp}
\begin{equation}
j_{l_{x},l_{y}}^{x}=it_{\mathrm{comp}}\left(P_{l_{x}+1,l_{y};l_{x},l_{y}}-P_{l_{x},l_{y};l_{x}+1,l_{y}}\right),\label{eq:jxcomp}
\end{equation}
\begin{equation}
j_{l_{x},l_{y}}^{y}=it_{\mathrm{comp}}\left(e^{il_{x}2\pi\alpha}P_{l_{x},l_{y}+1;l_{x},l_{y}}-e^{-il_{x}2\pi\alpha}P_{l_{x},l_{y};l_{x},l_{y}+1}\right).\label{eq:jycomp}
\end{equation}
\end {subequations} Since $P$ itself is not affected by the presence
of the Hamiltonian, the normalized current $\frac{j_{l_{x},l_{y}}^{\mu}}{t_{{\rm comp}}}$
is completely independent of $t_{{\rm comp}}$. As shown in Fig. \ref{fig:compPersist},
we find chiral currents strongly localized near the edges, much like
in the ground state of the equilibrium quantum Hall effect. We note
that the current in this compatible case is divergence-free, as expected
for a chiral edge mode. As a consequence, the dissipative current
$\dot{n}^{{\cal D}}\propto\nabla\cdot{\bf j}$ is zero, signaling
a zero local net flux of particles into or out of the baths.

\subsection {Incompatible hopping}\label{gaugeRef}

\begin{figure}
\begin{centering}
\includegraphics[bb=38.4149bp 0bp 686.186bp 300.96bp,scale=0.45]{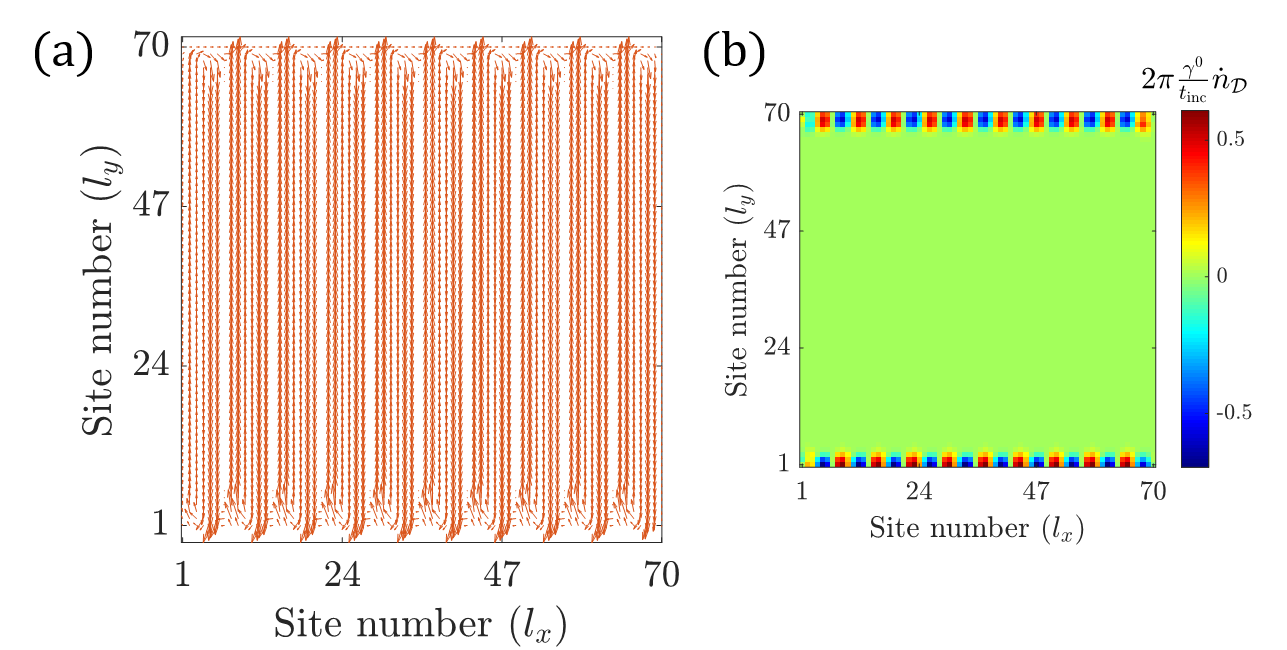}
\par\end{centering}
\caption{\label{fig:incPersist-Landau} (a) The current distribution calculated
in a square geometry in the incompatible-L scheme, Eqs. \eqref{eq:dissH}
and \eqref{eq:Hinc}. Each arrow is proportional to the current density
vector at the corresponding lattice site. (b) The normalized dissipative
current $\dot{n}_{\mathcal{D}}$, indicating the local exchange of
particles with the reservoir. Parameters used are $t_{{\rm inc}}=10^{-3}\gamma^{0}$,
$\alpha=\frac{1}{7}$ ($q=7$), $L_{x}=L_{y}=70$, and the optimal
values for $\mu^{*}$ and $\gamma^{{\rm in}}$.}
\end{figure}

In the case of the hopping Hamiltonian \eqref{eq:Hinc}, the current
is similarly given by

\begin {subequations}\label{jinc}
\begin{equation}
j_{l_{x},l_{y}}^{x}=it_{\mathrm{inc}}\left(P_{l_{x}+1,l_{y};l_{x},l_{y}}-P_{l_{x},l_{y};l_{x}+1,l_{y}}\right),\label{eq:jxinc}
\end{equation}
\begin{equation}
j_{l_{x},l_{y}}^{y}=it_{\mathrm{inc}}\left(P_{l_{x},l_{y}+1;l_{x},l_{y}}-P_{l_{x},l_{y};l_{x},l_{y}+1}\right).\label{eq:jyinc}
\end{equation}
\end {subequations} We find that the steady state currents which
develop in the system in such a scenario are completely different
than those of the compatible Hamiltonian, as we show in Fig. \ref{fig:incPersist-Landau}.
The current is not localized near the edges, but rather flows along
the $y$ direction with a structure periodic in $l_{x}$, determined
by the magnetic unit cell whose size is $q$ sites in this particular
dissipative configuration. The current trajectories terminate at the
system edges where they reverse their flow direction. This reversal
appears to be mediated by a local interchanging flux of particles
from or into the bath, as indicated by the non-vanishing dissipative
current $\dot{n}_{\mathcal{D}}$ near the edges. 

\begin{figure}
\begin{centering}
\includegraphics[bb=76.8539bp 0bp 684bp 321bp,scale=0.44]{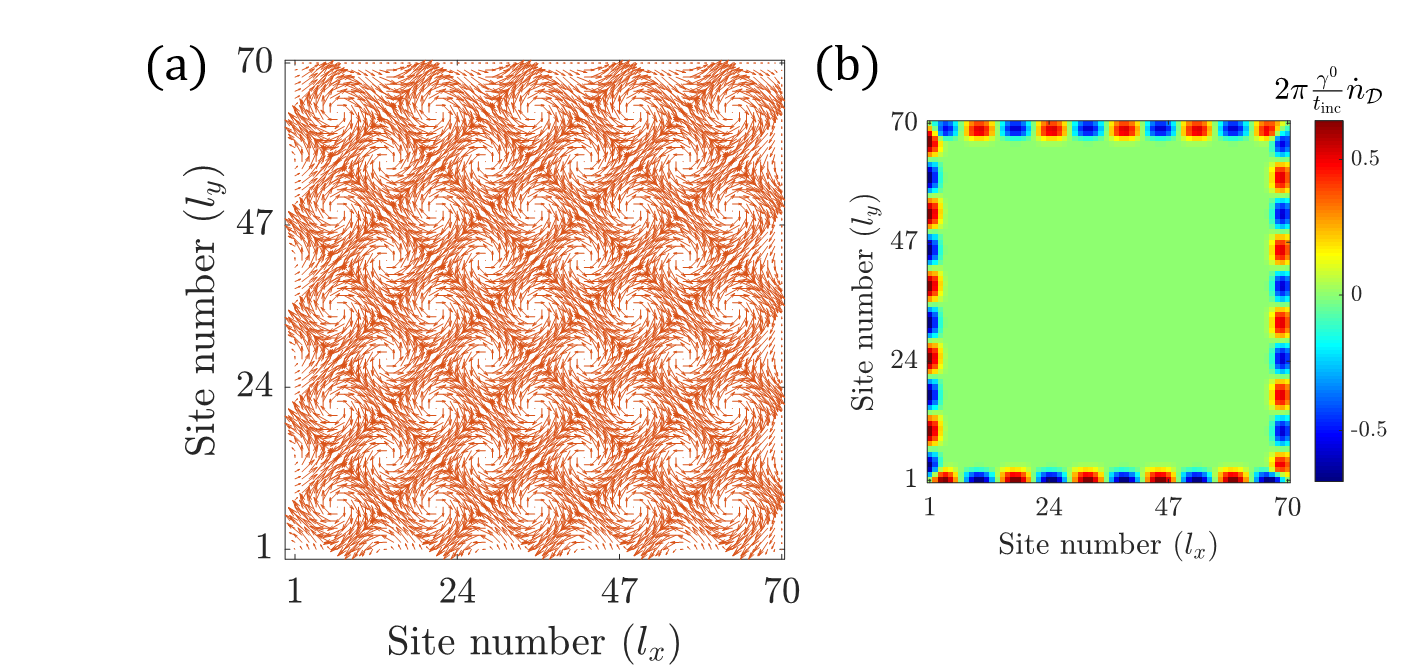}
\par\end{centering}
\caption{\label{fig:incPersist-Symm} (a) The current distribution calculated
in a square geometry in the incompatible-S scheme, Eqs. \eqref{eq:HdissS}
and \eqref{eq:Hinc}. Each arrow is proportional to the current density
vector at the corresponding lattice site. (b) The normalized dissipative
current $\dot{n}_{\mathcal{D}}$, indicating the local exchange of
particles with the reservoir. All parameters are the same as in Fig.
\ref{fig:incPersist-Landau}.}
\end{figure}

This result seems to be somewhat peculiar at first, since the current
in the system appears to prefer one direction ($y$) over the other
($x$). This is a result of using, in the incompatible L-scheme, the
dissipative Hamiltonian \eqref{eq:dissH}, with its anisotropic phase
factors, in combination with the isotropic incompatible system Hamiltonian
\eqref{eq:Hinc}. This is not the case in the gauge-inequivalent incompatible
S-scheme, Eq. \eqref{eq:HdissS} and \eqref{eq:Hinc}. And indeed,
in that case, as evident in Fig. \ref{fig:incPersist-Symm}, the current
pattern which emerges has circulating currents around plaquettes with
$q\times q$ sites (recall that $\alpha=\frac{p}{q}$, so for, e.g.,
$\alpha=\frac{1}{7}$ one has $q=7$).

Qualitatively, the current patterns do not change appreciably as $t_{{\rm inc}}$
increases beyond the $t_{{\rm inc}}\ll\gamma^{0}$ limit. What we
observe is a steep decline in the amplitude of the normalized current
$\frac{j}{t_{{\rm inc}}}$ as $t_{{\rm inc}}$ becomes comparable
to the dissipative scale $\gamma^{0}$. This is understood by the
depletion of current carriers in the system as the steady state occupation
deteriorates, and is compensated by sufficiently increasing the value
of the refilling rate $\gamma^{{\rm in}}$.

\section {conductance}\label{conductivity}

One of the most remarkable properties of the quantum Hall ground state
is its exactly quantized transverse electrical conductance. In order
to reveal the response and transport properties of the dissipatively
prepared state, we introduce a small perturbation to the Hamiltonian,
playing the role of an electric field along the $x$ direction. Using
the relation between the electric field ${\bf E}$ and the electrostatic
potential $\phi$, $\mathbf{E}=-\nabla\phi$, we incorporate a perturbative
term synonymous with $\int d\mathbf{x}\rho\left(x\right)\phi\left(x\right)$,

\noindent 
\begin{equation}
\delta H_{{\bf E}}=-\sum_{l_{x},l_{y}}\left(E_{x}l_{x}+E_{y}l_{y}\right)a_{l_{x},l_{y}}^{\dagger}a_{l_{x},l_{y}},\label{eq:deltH-1}
\end{equation}
with the electric field ${\bf E}=(E_{x},E_{y},0)$. As we are only
interested in the linear response regime, we calculate the first-order
change in the occupation numbers matrix $\delta P$ due to a finite
value of $\mathbf{E}$, which amounts to solving another matrix Sylvester
equation,
\begin{equation}
\left[\frac{\Gamma^{\mathrm{in}}+\Gamma^{\mathrm{out}}}{2}+ih\right]\delta P+\delta P\left[\frac{\Gamma^{\mathrm{in}}+\Gamma^{\mathrm{out}}}{2}-ih\right]=i\left[P_{0},\delta h\right],\label{eq:sylvesterLRT}
\end{equation}
with $P_{0}$ the solution of Eq. \eqref{eq:Psylvester-1} without
an applied field, and $\delta h$ is the single particle matrix corresponding
to $\delta H$. To the best of our knowledge, a Sylvester/Lyapunov
equation of the kind of Eq. \eqref{eq:sylvesterLRT} has not been
considered before as a method for calculating linear response properties.
We calculate the modification of the values of the current due to
this perturbation, $\delta j_{l_{x},l_{y}}^{\mu}=j_{l_{x},l_{y}}^{\mu}\left(P\rightarrow P+\delta P\right)-j_{l_{x},l_{y}}^{\mu}\left(P\right)$,
and find the conductance tensor, which we define by

\noindent \begin {subequations}\label{eq:condtensor} 
\begin{equation}
G_{xx}=\lim_{E_{x}\to0}\frac{1}{E_{x}}\sum_{l_{y}}\frac{1}{L_{y}}\delta j_{\frac{1}{2}L_{x},l_{y}}^{x},
\end{equation}
\begin{equation}
G_{yx}=\lim_{E_{x}\to0}\frac{1}{E_{x}}\sum_{l_{x},l_{y}}\frac{1}{L_{y}L_{x}}\delta j_{l_{x},l_{y}}^{y},
\end{equation}
with $E_{y}=0$, and
\begin{equation}
G_{yy}=\lim_{E_{y}\to0}\frac{1}{E_{y}}\sum_{l_{x}}\frac{1}{L_{x}}\delta j_{l_{x},\frac{1}{2}L_{y}}^{y},
\end{equation}
\begin{equation}
G_{xy}=\lim_{E_{y}\to0}\frac{1}{E_{y}}\sum_{l_{x},l_{y}}\frac{1}{L_{x}L_{y}}\delta j_{l_{x},l_{y}}^{x},
\end{equation}
\end {subequations} with $E_{x}=0$. Notice that in all terms of
$G$ we average over the dimension perpendicular to the electric field
applied. In all calculations presented below we always have periodic
boundary conditions in that dimension. For more details and the justification
of Eq. \eqref{eq:condtensor}, see Appendix \eqref{appC}. Similarly
to the previous section, we discuss our findings separately for the
two possible Hamiltonian classes.

\subsection {Compatible hopping}

\begin{figure}
\begin{centering}
\includegraphics[bb=14.39126bp 0bp 472.033bp 385.037bp,scale=0.57]{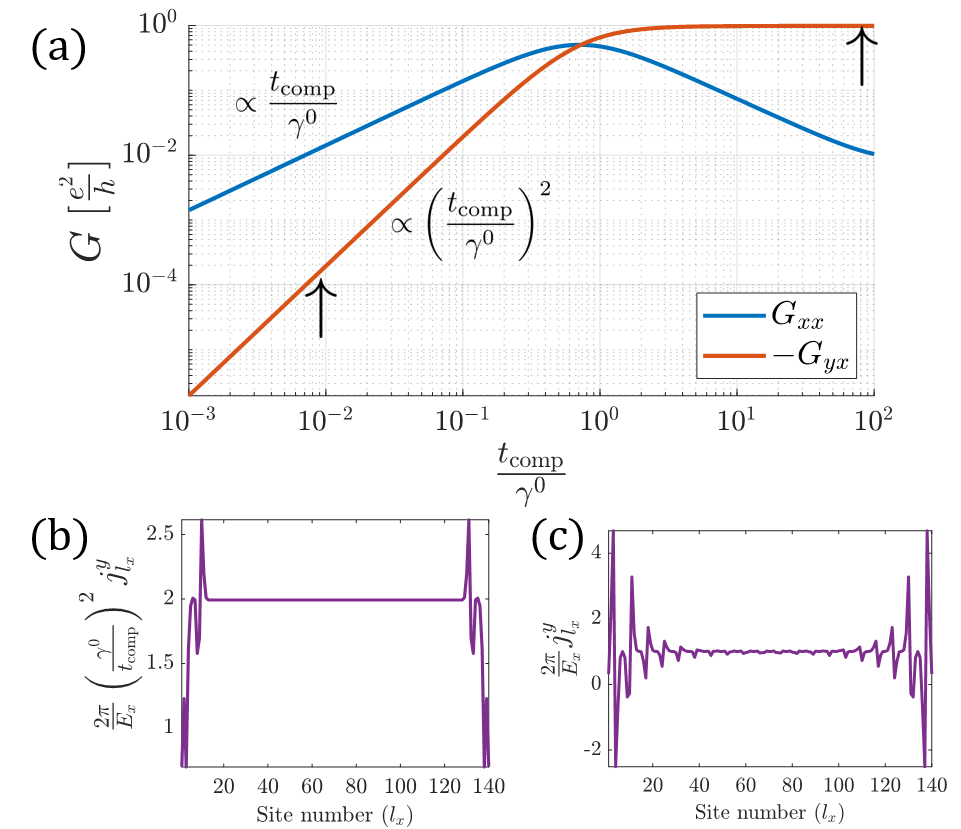}
\par\end{centering}
\caption{\label{fig:condRegimes_compatible} (a) conductance calculations for
the case of a compatible system Hamiltonian. $G_{xx}$ (blue) and
$G_{yx}$ (red) are displayed as a function of the ratio $\frac{t_{{\rm comp}}}{\gamma^{0}}$.
The behavior in the dissipative regime is indicated, whereas in the
Hamiltonian regime the transverse conductance is quantized to $G_{yx}\approx-\frac{e^{2}}{h}$,
the value of the topological invariant, the Chern number. (b-c) Representative
transverse current patterns of $j_{l_{x}}^{y}=\sum_{l_{y}}j_{l_{x},l_{y}}^{y}$
along the systems, for (b) $t_{{\rm comp}}=0.01\gamma^{0}$ and (c)
$t_{{\rm comp}}=100\gamma^{0}$ {[}the corresponding values are marked
by arrows in (a){]}. Throughout this figure we used a cylindrical
geometry with edges along the $\hat{x}$ direction with $L_{x}=L_{y}=140$,
$\alpha=\frac{1}{7}$, and the optimal values for $\mu^{*}$ and $\gamma^{{\rm in}}$.}
\end{figure}

Unlike our discussion regarding the persistent currents, it is clear
from Eq. \eqref{eq:sylvesterLRT} that even in the compatible case,
the relative amplitude of the hopping Hamiltonian, as compared to
dissipation rate, may play an important role. We indeed find two regimes
for the conductance in our system, corresponding to $\frac{t_{{\rm comp}}}{\gamma^{0}}\gg1$
and $\frac{t_{{\rm comp}}}{\gamma^{0}}\ll1$ with a smooth crossover
around $t_{{\rm comp}}\approx\gamma^{0}$, see Fig. \ref{fig:condRegimes_compatible}a.
Interestingly, in both regimes the transverse currents flows near
system edges in a similar fashion (Fig. \ref{fig:condRegimes_compatible}b,c).

In the large $\frac{t_{{\rm comp}}}{\gamma^{0}}$ case, which we refer
to as the Hamiltonian regime, we observe quantum-Hall-like behavior,
namely the quantization of the transverse response in accordance with
the topological Chern number associated with the steady-state density
matrix, which is equal to the one associated with the lowest band
of $H^{\mathrm{diss}}$ \citep{MG}. The value at the conductance
plateau stabilizes roughly when $t_{{\rm comp}}$is larger than $10\gamma^{0}$.
For the parameters used in Fig. \ref{fig:condRegimes_compatible},
the value of the Hall conductance at the plateau is $G_{yx}=-0.984\frac{e^{2}}{h}$,
and we have checked numerically that the deviation from ideal quantization
vanishes at the thermodynamic limit and scales as $\propto\frac{1}{L_{x}}$.
Alongside this approximate quantization, we observe a vanishing longitudinal
response with increasing $t_{{\rm comp}}$, as expected from the limit
$t_{{\rm comp}}\gg\gamma^{0}$, where an insulating behavior is predicted.
The correction to this asymptotic limit seems to scale approximately
with $\frac{\gamma^{0}}{t_{{\rm comp}}}$, indicating that the first
order correction in the dissipation to Eq. \eqref{eq:sylvesterLRT}
is dominant and does not cancel out.

In the opposite limit, the dissipative regime, both transverse and
longitudinal responses are small, and increase as $t_{{\rm comp}}$
becomes larger. While $G_{xx}$ grows linearly with $\frac{t_{{\rm comp}}}{\gamma^{0}}$,
the transverse conductance $G_{yx}$ is proportional to $\left(\frac{t_{{\rm comp}}}{\gamma^{0}}\right)^{2}$.
To understand why the linear contribution to $G_{yx}$ vanishes, although
the current  itself is proportional to $t_{{\rm comp}}$, one must
examine the $h=0$ limit of Eq. \eqref{eq:sylvesterLRT}. Performing
a Fourier transform of all the fermionic operators in the system with
respect to the $y$ direction, $a_{l_{x},l_{y}}\sim\sum_{k_{y}}e^{ik_{y}l_{y}}a_{l_{x},k_{y}}$,
one finds that in this basis the matrices $\left(\Gamma^{\mathrm{in}}+\Gamma^{\mathrm{out}}\right)$,
$P_{0}$, and $\delta h$ are purely real, which would make $\delta P$
in this limit purely imaginary. Since $\delta P$ is hermitian, its
diagonal elements $\left\langle a_{l_{x},k_{y}}^{\dagger}a_{l_{x},k_{y}}\right\rangle $
vanish. But $j^{y}$ is composed of exactly these vanishing averages.
Hence $G_{yx}$ is zero to first order in $\frac{t_{{\rm comp}}}{\gamma^{0}}$.

\begin{figure}
\begin{centering}
\includegraphics[bb=38.4067bp 0bp 501.208bp 225.1261bp,scale=0.6]{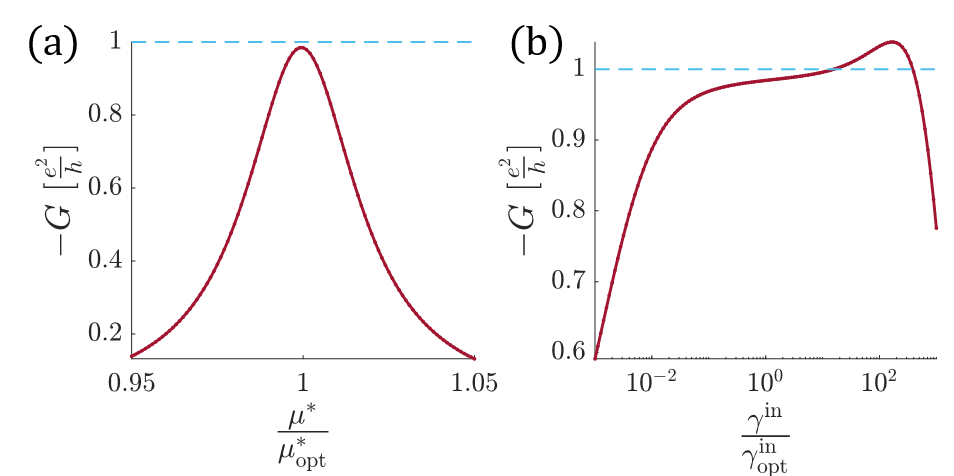}
\par\end{centering}
\caption{\label{fig:condRegimes_compatibleTuning} Parameter dependence of
the transverse conductance $G_{yx}$ in the Hamiltonian regime of
the compatible hopping scenario. (a) $\mu^{*}$ is changed and $\gamma^{{\rm in}}=\gamma_{{\rm opt}}^{{\rm in}}$
is kept constant. (b) $\gamma^{{\rm in}}$ is changed and $\mu^{*}=\mu_{{\rm opt}}^{*}$
is kept constant. The dashed light blue line marks the quantized value
$G_{yx}=-1$. Throughout this figure we used a cylindrical geometry
with edges along the $x$ direction with $L_{x}=L_{y}=140$, $\alpha=\frac{1}{7}$,
and $t_{{\rm comp}}=100\gamma^{0}$.}
\end{figure}

We have also investigated how the quantized conductance (in the Hamiltonian
regime) changes when the parameters controlling the dissipative scheme,
i.e., $\mu^{*}$ and $\gamma^{{\rm in}}$, are modified, and do not
assume their optimal values. We find that a precise tuning of $\mu^{*}$
is required for the nearly precise quantization (Fig. \ref{fig:condRegimes_compatibleTuning}a),
and that a small deviation from the optimal value deteriorates $G_{yx}$
significantly. This is because $\mu^{*}$ should be tuned to minimize
$\gamma_{{\bf k},1}^{{\rm out}}$, such that the $\lambda=1$ band
would be nearly filled in the steady state. Once $\mu^{*}$ is detuned
away from the middle of the first band, this band will always be partially
filled (unless we increase $\gamma^{{\rm in}}$ to the point that
higher bands are no longer nearly empty). In contrast, the transverse
response is somewhat less sensitive to changes in $\gamma^{{\rm in}}$,
and we find that only changes of order of magnitude have an appreciable
effect (Fig. \ref{fig:condRegimes_compatibleTuning}b): A small change
of the refilling rate does little to change the occupation of the
different bands, leading to small deviations from the quantized value.

\subsection {Incompatible hopping}

\begin{figure}
\begin{centering}
\includegraphics[bb=0bp 0bp 399.36bp 271.7332bp,scale=0.65]{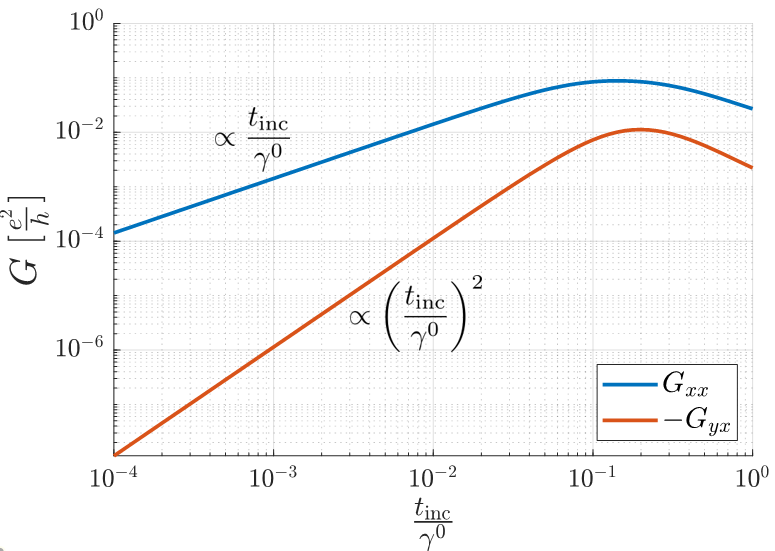}
\par\end{centering}
\caption{\label{fig:condRegimes_incompatible} The conductance calculated in
the case of an incompatible-L scheme, Eqs. \eqref{eq:dissH} and \eqref{eq:Hinc}.
$G_{xx}$ (blue) and $G_{yx}$ (red) are displayed as a function of
the ratio $\frac{t_{{\rm inc}}}{\gamma^{0}}$. The behavior at small
$\frac{t_{{\rm inc}}}{\gamma^{0}}$ is indicated. We used a cylindrical
geometry with edges along the $x$ direction with $L_{x}=L_{y}=140$,
$\alpha=\frac{1}{7}$, and the optimal values for $\mu^{*}$ and $\gamma^{{\rm in}}$.}
\end{figure}

Similarly to what we have seen for the steady state persistent currents,
things change when one has an Hamiltonian incompatible with the magnetic
flux in \eqref{eq:dissH}. Whereas in the dissipative regime with
$t_{{\rm inc}}\ll\gamma^{0}$ the conductance features similar dependence
on $\frac{t_{{\rm inc}}}{\gamma^{0}}$ as in the coherent case, at
higher $\frac{t_{{\rm inc}}}{\gamma^{0}}$ the conductance begins
to decline in amplitude, never reaching the topological quantized
value for $G_{yx}$, see Fig. \ref{fig:condRegimes_incompatible}.
Again, this should not be surprising, as we have already seen that
a large $t_{{\rm inc}}$ negates our ability to manipulate the particles
into the desired dissipatively engineered steady state, which possesses
some QHE-like features. Moreover, increasing $t_{{\rm inc}}$ beyond
the dissipative regime without altering $\gamma^{{\rm in}}$, we in
fact get a much less populated steady state, which in turn can carry
significantly less current. We note that in the dissipative regime,
the current distribution is also indistinguishable from the behavior
for the compatible Hamiltonian, e.g., Fig. \ref{fig:condRegimes_compatible}b.

\begin{figure}
\begin{centering}
\includegraphics[bb=38.3746bp 0bp 529.09bp 356.636bp,scale=0.53]{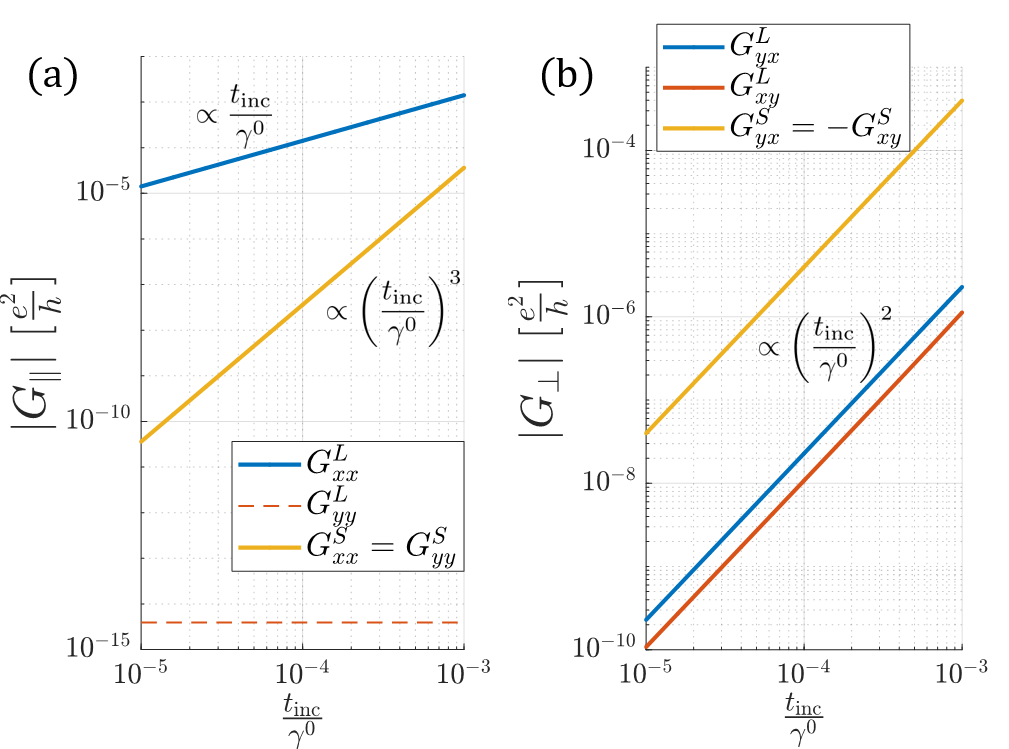}
\par\end{centering}
\caption{\label{fig:anisotropicGinc} Different elements of the conductance
matrix, calculated for the incompatible Hamiltonian in the dissipative
regime for the incompatible-L and incompatible-S schemes. The conductance
for both directions in the ``L'' choice (red and blue), and for the
symmetric ``S'' choice (yellow) is presented for (a) the longitudinal
and (b) the transverse elements. For the symmetric choice we present
only one element of each, since the remaining ones are identical.
$G_{yy}^{L}$ is found to be zero up to numerical noise, which is
at the level depicted by the corresponding dashed trace. The behavior
as a function of $\frac{t_{{\rm inc}}}{\gamma^{0}}$ is indicated
next to each plot. The calculations are performed with edges in the
direction perpendicular to the applied field, with $L_{x}=L_{y}=70$,
$\alpha=\frac{1}{7}$, and the optimal values for $\mu^{*}$ and $\gamma^{{\rm in}}$.}
\end{figure}

Lastly, we find that the conductance matrix is anisotropic, namely
that $G_{xx}\neq G_{yy}$ and $G_{yx}\neq-G_{xy}$ in this incompatible
dissipative regime, which is not the case for the compatible Hamiltonian.
Once more, this is nothing but a consequence of the anisotropy of
the dissipation Hamiltonian \eqref{eq:dissH}, which explicitly has
a preferable axis, when combined with the incompatible system Hamiltonian
\eqref{eq:Hinc}. Using instead the incompatible-S scheme, Eqs. \eqref{eq:HdissS}
and \eqref{eq:Hinc}, removes this anisotropy, as one would expect.
In Fig. \ref{fig:anisotropicGinc} we show calculation of different
elements in the conductance matrix for both cases. For the longitudinal
conductance, i.e., $G_{xx}$ and $G_{yy}$, we find not only different
values, but also different functional form, which depends on both
the bath coupling convention and the direction. In the incompatible-L
scheme, $G_{yy}$ appears to vanish (up to numerical noise). This
is due to the oscillatory nature of the underlying steady state along
the $\hat{x}$ direction, as implied by the persistent currents in
Fig. \ref{fig:incPersist-Landau}. There is nearly perfect cancellation
of counterflowing current within the magnetic unit cell, giving rise
to a vanishing total response to the electric field perturbation.
Similar oscillatory behavior apparent from Fig. \ref{fig:incPersist-Symm}
also explains the much weaker longitudinal response that scales as
$\propto\left(\frac{t_{{\rm inc}}}{\gamma^{0}}\right)^{3}$in the
more symmetric incompatible-S scheme. It appears that the finite response
in this case stems from the violation of perfect periodicity of the
the persistent current flow by the edges of the cylinder, see again
Fig. \ref{fig:incPersist-Symm}. This violation does not exist in
the asymmetric incompatible-L scheme since according to Fig. \ref{fig:incPersist-Landau},
the edges of the cylinder are parallel to the persistent current flow
and do not affect it. The transverse response however, shows a more
universal trend of $\propto\left(\frac{t_{{\rm inc}}}{\gamma^{0}}\right)^{2}$,
albeit with different amplitudes for $G_{xy}$, $G_{yx}$ in the asymmetric
case and for $G_{yx}=-G_{xy}$ in the symmetric one. Again, it appears
that the reason for the L/S difference, also in the transverse conductance,
is the susceptibility of the underlying steady state to current flow,
which can be inferred from the persistent current patterns, Figs.
\ref{fig:incPersist-Landau} and \ref{fig:incPersist-Symm}.

\section {Conclusions}\label{conclusions}

We investigate the transport properties of the purely dissipative
theoretical scheme, presented in Ref. \citep{MG}, which reproduces
a state synonymous with a very low temperature equilibrium quantum
Hall state in a 2D lattice exploiting engineered dissipation. These
properties can not be probed without introducing some coherent Hamiltonian
dynamics for the lattice particles. We have demonstrated that a departure
from the completely dissipative scheme, crucial for the transport
study, as well as being more experimentally realistic, does not necessarily
harm the engineered steady state, provided that the Hamiltonian is
compatible with the evaporative part, or alternatively, that the incompatible
part of that Hamiltonian is sufficiently small in magnitude compared
to the dissipative energy scale.

Having introduced the Hamiltonian dynamics, we could explore the persistent
currents that flow once the dissipative steady state is reached in
the system. In the case where the lattice Hamiltonian has the same
magnetic field present in the evaporative dynamics the main observation
is the well-known chiral edge-modes, characteristic of equilibrium
quantum Hall and Chern insulator systems, and the lack of any other
currents in the system. On the other hand, a hopping Hamiltonian lacking
the compatible magnetic field, induces bulk currents in a pattern
and a direction determined by the chosen pattern of phases of the
artificial gauge field (the gauge inequivalent L vs. S schemes). Near
the edges the net flow to/from the reservoirs becomes finite locally
(though it sums up to zero globally), facilitating current backscattering
at the edges via the particle reservoir.

The electrical conductance of the dissipative steady state was also
examined. Quantization of the transverse conductance, consistent with
the theoretical Chern number, was shown to arise for the compatible
Hamiltonian, provided it is sufficiently larger as compared to the
dissipative rates. The quality of this quantization is affected by
the accuracy of the assigned dissipative parameters, namely the refilling
rate $\gamma^{\mathrm{in}}$ and $\mu^{*}$, which allows to tune
the evaporation rate of the filled band. We find the quantization
is much less sensitive to the former compared to the latter. A weaker
compatible Hamiltonian featured mainly currents in the vicinity of
the edges mediated by a dissipative current, together with a weak
second order transverse response, also located near the edges of the
system. The conductance matrix in the incompatible regime is slightly
more complicated in structure, as the choice of system-bath coupling
details (incompatible-L vs. incompatible-S schemes) may render it
anisotropic: we observe longitudinal response in the $x$ direction
identical to the one obtained with a weak compatible Hamiltonian,
and different power law trends depending on the scheme. The transverse
response seems to have a universal $\propto t_{{\rm inc}}^{2}$ behavior,
but with anisotropic and scheme dependent absolute values.

Under the circumstances of a compatible hopping Hamiltonian, sufficiently
larger in amplitude than the dissipative energy scale, the potential
for obtaining some of the equilibrium quantum Hall properties, such
as chiral edge-states and quantized conductance, may be put to the
test by checking its robustness to some finite amount of disorder,
both in the bath interaction and lattice Hamiltonians, which will
be the subject of a future study. The way the transverse current is
carried through the system in its dissipative steady state, under
such perturbations, can be probed using the tools we have developed,
and may shed further insights into the evaporative processes occurring
in the system. This also opens up the possibility for engineering
increasingly complex coupling to the bath, allowing exploration of
more interesting possibilities and topological traits through the
engineering of dissipative two-particle interactions. This in turn
would allow the exploration of the dissipative analogues of exotic
fractional states \citep{fractional0,Fractional1}, and perhaps the
equivalents of anyons \citep{anyons1,anyons2,QuantComp}.

\section {Acknowledgments} 

We would like to thank B. A. Bernevig, B. Bradlyn, J. Budich, J. I.
Cirac, N. Cooper, S. Diehl, M. Hafezi, Y. Gefen, A. Gorshkov, A. Kemenev,
N. Schuch, H. H. Tu, and P. Zoller for useful discussions. This work
has been supported by the Israel Science Foundation (Grant No. 227/15),
the German Israeli Foundation (Grant No. I-1259-303.10), the US-Israel
Binational Science Foundation (Grants No. 2014262 and 2016224), and
the Israel Ministry of Science and Technology (Contract No. 3-12419).

\begin {appendix}

\section {Steady state occupation numbers}\label{SSoccP}

In this appendix, we re-derive some of the results of Ref. \citep{Lindbladians},
and bring the parts relevant to our discussion for the reader's convenience.
We begin by introducing a quadratic Hamiltonian for the lattice particles
$a$ (which in this work will be either the compatible Hamiltonian
\eqref{eq:Hcomp}, or the incompatible Hamiltonian \eqref{eq:Hinc}),

\begin{equation}
H=\sum_{A,B}h_{AB}a_{A}^{\dagger}a_{B},
\end{equation}

\noindent with generalized indexes $A,B,$ which may each represent
for example two spatial indexes (e.g., $A=l_{x},l_{y}$ and $B=l_{x}+1,l_{y}-1$).

\noindent The dissipator is also assumed to be quadratic, so that
the Lindbladian master equation for the density matrix is
\begin{align}
\frac{d}{dt}\rho & =-i\sum_{A,B}h_{AB}\left[a_{A}^{\dagger}a_{B},\rho\right]+\nonumber \\
 & \sum_{A,B}\Gamma_{AB}^{\left(1\right)}\left(a_{B}\rho a_{A}^{\dagger}-\frac{1}{2}\left\{ a_{A}^{\dagger}a_{B},\rho\right\} \right)+\nonumber \\
 & \sum_{A,B}\Gamma_{AB}^{\left(2\right)}\left(a_{A}^{\dagger}\rho a_{B}-\frac{1}{2}\left\{ a_{B}a_{A}^{\dagger},\rho\right\} \right),\label{eq:GeneralME}
\end{align}

\noindent with the $\Gamma$ matrices encapsulating the dissipative
processes. In this work, we have the matrices
\begin{equation}
\Gamma^{\left(1\right)}=\Gamma^{\mathrm{out}}=2\frac{2\pi}{\hbar}\nu_{0}H_{{\rm diss}}^{2},\label{eq:Gamma1Gammout}
\end{equation}
\begin{equation}
\Gamma^{\left(2\right)}=\Gamma^{{\rm in}}.
\end{equation}
We define the following matrix of expectation values $P_{AB}\left(t\right)=\left\langle a_{A}^{\dagger}a_{B}\right\rangle _{t}$,
with the shorthand notation $\left\langle M\right\rangle _{t}=\mathrm{Tr}\left\{ \rho\left(t\right)M\right\} $.
According to Eq. \eqref{eq:GeneralME}, the dynamics of this matrix
is given by 
\begin{align}
\frac{d}{dt}P_{CD}\left(t\right) & =-i\sum_{A,B}h_{AB}\left\langle \left[a_{C}^{\dagger}a_{D},a_{A}^{\dagger}a_{B}\right]\right\rangle _{t}\nonumber \\
 & +\sum_{A,B}\Gamma_{AB}^{\left(1\right)}\left\langle a_{A}^{\dagger}a_{C}^{\dagger}a_{D}a_{B}-\left\{ a_{A}^{\dagger}a_{B},\frac{a_{C}^{\dagger}a_{D}}{2}\right\} \right\rangle _{t}\nonumber \\
 & +\sum_{A,B}\Gamma_{AB}^{\left(2\right)}\left\langle a_{B}a_{C}^{\dagger}a_{D}a_{A}^{\dagger}-\left\{ a_{B}a_{A}^{\dagger},\frac{a_{C}^{\dagger}a_{D}}{2}\right\} \right\rangle _{t}.
\end{align}

\noindent Using the definition for $P_{AB}\left(t\right)$, Wick's
theorem, and the fermionic anti-commutation relations, we find the
matrix equation 
\begin{equation}
\frac{d}{dt}P\left(t\right)=-i\left[h,P\left(t\right)\right]-\frac{1}{2}\left\{ \Gamma^{\left(1\right)}+\Gamma^{\left(2\right)},P\left(t\right)\right\} +\Gamma^{\left(2\right)}.\label{eq:Pdot}
\end{equation}

\noindent The steady state version of Eq. \eqref{eq:Pdot} can be
manipulated into a Sylvester equation for the matrix $P$, Eq. \eqref{eq:Psylvester-1}
of the main text. If $\Gamma^{\left(1\right)}$ and $\Gamma^{\left(2\right)}$
can be simultaneously diagonalized, such as in our case, where $\Gamma^{\left(2\right)}$
is a constant times unity matrix, the steady solution for \eqref{eq:Psylvester-1}
in the purely dissipative regime ($h\rightarrow0$) is given (in the
basis where the $\Gamma$ matrices are diagonal) by
\begin{equation}
P=\frac{\Gamma^{\left(2\right)}}{\Gamma^{\left(1\right)}+\Gamma^{\left(2\right)}}.\label{eq:compP}
\end{equation}

\noindent This reduces to Eq. \eqref{eq:occDiss} in the main text.
Notice however, that this result remains intact even in the presence
of an Hamiltonian $h$ which is diagonal in the same basis as the
$\Gamma$ matrices, due to the cancellation of the commutator term
in \eqref{eq:Pdot}. This is the reason for the distinction between
the compatible and the incompatible Hamiltonians: In the former case
\eqref{eq:compP} holds, but not in the latter.

\section {The current operators}\label{currentApp}

Consider the expectation value of the particle density in the site
$\left(l_{x},l_{y}\right)$ of the lattice, $n_{l_{x},l_{y}}\left(t\right)\equiv\left\langle \rho\left(t\right)a_{l_{x},l_{y}}^{\dagger}a_{l_{x},l_{y}}\right\rangle =P_{l_{x},l_{y};l_{x},l_{y}}\left(t\right)$.
Its time-evolution is given by 
\begin{align}
\frac{d}{dt}n_{l_{x},l_{y}}\left(t\right) & =\dot{n}_{l_{x},l_{y}}^{H}\left(t\right)+\dot{n}_{l_{x},l_{y}}^{\mathcal{D}}\left(t\right),
\end{align}
with
\begin{align}
\dot{n}_{l_{x},l_{y}}^{H}\left(t\right) & =-i\sum_{n_{x},n_{y}}h_{l_{x},l_{y};n_{x},n_{y}}\left\langle a_{n_{x},n_{y}}^{\dagger}a_{l_{x},l_{y}}\right\rangle _{t}\nonumber \\
 & +i\sum_{n_{x},n_{y}}\left\langle a_{l_{x},l_{y}}^{\dagger}a_{n_{x},n_{y}}\right\rangle _{t}h_{n_{x},n_{y};l_{x},l_{y}},
\end{align}
\begin{align}
\dot{n}_{l_{x},l_{y}}^{\mathcal{D}} & \left(t\right)=-\frac{1}{2}\sum_{n_{x},n_{y}}\left[\Gamma^{\left(1\right)}+\Gamma^{\left(2\right)}\right]_{l_{x},l_{y};n_{x},n_{y}}\left\langle a_{n_{x},n_{y}}^{\dagger}a_{l_{x},l_{y}}\right\rangle _{t}\nonumber \\
 & -\frac{1}{2}\sum_{n_{x},n_{y}}\left\langle a_{l_{x},l_{y}}^{\dagger}a_{n_{x},n_{y}}\right\rangle _{t}\left[\Gamma^{\left(1\right)}+\Gamma^{\left(2\right)}\right]_{l_{x},l_{y};n_{x},n_{y}}\nonumber \\
 & +\Gamma_{l_{x},l_{y};l_{x},l_{y}}^{\left(2\right)}.
\end{align}

\noindent Plugging in $H_{\mathrm{comp}}$, we get
\begin{align}
\dot{n}_{l_{x},l_{y}}^{H}\left(t\right) & =-it_{\mathrm{comp}}\left\langle a_{l_{x}+1,l_{y}}^{\dagger}a_{l_{x},l_{y}}\right\rangle _{t}\nonumber \\
 & +it_{\mathrm{comp}}\left\langle a_{l_{x},l_{y}}^{\dagger}a_{l_{x}-1,l_{y}}\right\rangle _{t}\nonumber \\
 & -it_{\mathrm{comp}}e^{il_{x}2\pi\alpha}\left\langle a_{l_{x},l_{y}+1}^{\dagger}a_{l_{x},l_{y}}\right\rangle _{t}\nonumber \\
 & +it_{\mathrm{comp}}e^{il_{x}2\pi\alpha}\left\langle a_{l_{x},l_{y}}^{\dagger}a_{l_{x},l_{y}-1}\right\rangle _{t}+{\rm h.c.\,}.
\end{align}

\noindent The r.h.s. can be written as a discrete divergence,
\begin{equation}
\dot{n}_{l_{x},l_{y}}^{H}\left(t\right)=-\left(j_{l_{x}+1,l_{y}}^{x}-j_{l_{x},l_{y}}^{x}\right)-\left(j_{l_{x},l_{y}+1}^{y}-j_{l_{x},l_{y}}^{y}\right),\label{eq:ndotH}
\end{equation}
which allows us to define the currents as in Eq. \eqref{jcomp}. A
transformation $t_{{\rm comp}}\rightarrow t_{{\rm inc}}$, $\alpha\rightarrow0$,
then gives the current  in the presence of $H_{{\rm inc}}$, Eq. \eqref{jinc}.

A closer look at $\dot{n}^{\mathcal{D}}$ reveals that it can be separated
into two terms, corresponding, respectively, to flow of particles
from the refilling reservoir into the system, and particles leaving
the system into the evaporative reservoir,

\begin{align}
\dot{n}_{l_{x},l_{y}}^{\mathcal{D}} & \left(t\right)=\frac{1}{2}\left\{ \Gamma^{\left(2\right)},1-P\right\} _{l_{x},l_{y};l_{x},l_{y}}-\frac{1}{2}\left\{ \Gamma^{\left(1\right)},P\right\} _{l_{x},l_{y};l_{x},l_{y}}\nonumber \\
 & \equiv J_{l_{x},l_{y}}^{\mathcal{D},\mathrm{in}}-J_{l_{x},l_{y}}^{\mathcal{D},\mathrm{out}}.\label{eq:ndotbrokenapart}
\end{align}
Also note that for a compatible Hamiltonian, plugging in the steady
state $P$, given by Eq. \eqref{eq:compP}, leads to the steady state
$\dot{n}^{\mathcal{D}}$ vanishing exactly. More generally, in the
steady state $\frac{d}{dt}n=0$, hence the ``dissipative current''
is immediately found (for any choice of Hamiltonian) to be
\begin{equation}
\dot{n}_{l_{x},l_{y}}^{\mathcal{D}}=-\dot{n}_{l_{x},l_{y}}^{H},\label{eq:dissipativeDiv}
\end{equation}
with the r.h.s. given by Eq. \eqref{eq:ndotH}.

For completeness, we also list here the expressions for the current
operators for all choices of $H^{{\rm sys}}$. With the system Hamiltonian
$H_{{\rm comp},S}^{{\rm sys}}$ given by Eq. \eqref{eq:HcompS}, we
find
\begin{equation}
j_{l_{x},l_{y};{\rm comp},S}^{x}=it_{\mathrm{comp}}\left(e^{-il_{y}\pi\alpha}P_{l_{x}+1,l_{y};l_{x},l_{y}}-e^{il_{y}\pi\alpha}P_{l_{x},l_{y};l_{x}+1,l_{y}}\right),\label{eq:jxcomp-2}
\end{equation}
\begin{equation}
j_{l_{x},l_{y};{\rm comp},S}^{y}=it_{\mathrm{comp}}\left(e^{il_{x}\pi\alpha}P_{l_{x},l_{y}+1;l_{x},l_{y}}-e^{-il_{x}\pi\alpha}P_{l_{x},l_{y};l_{x},l_{y}+1}\right).\label{eq:jycomp-2}
\end{equation}
For the choice of $\tilde{H}_{{\rm inc}}^{{\rm sys}}$ in Eq. \eqref{eq:HincTransformed},
the current operators are given by
\begin{equation}
j_{l_{x},l_{y};\tilde{{\rm inc}}}^{x}=it_{\mathrm{inc}}\left(e^{-il_{y}\pi\alpha}P_{l_{x}+1,l_{y};l_{x},l_{y}}-e^{il_{y}\pi\alpha}P_{l_{x},l_{y};l_{x}+1,l_{y}}\right),\label{eq:jxinc-1}
\end{equation}
\begin{equation}
j_{l_{x},l_{y};\tilde{{\rm inc}}}^{y}=it_{\mathrm{inc}}\left(e^{-il_{x}\pi\alpha}P_{l_{x},l_{y}+1;l_{x},l_{y}}-e^{il_{x}\pi\alpha}P_{l_{x},l_{y};l_{x},l_{y}+1}\right).\label{eq:jyinc-1}
\end{equation}
Of course, for choices related by a physical gauge transformation,
although the expression for the current may look different, its expectation
value is gauge independent. On the other hand, we note that whether
we work with the gauge-inequivalent incompatible-L or incompatible-S
choices, the expressions for the current operators, Eqs. \eqref{eq:jxinc}--\eqref{eq:jyinc}
are unaltered. However, since $H^{{\rm diss}}$ differs, the steady-state
$P$ matrix will vary, and thus the calculated currents. 

An alternative way to derive the form of the different current operators
on the lattice is outlined below. One begins with the total Hamiltonian,
comprise of $a$-particles hopping terms, $a$-$b$ interactions,
and the bath particles Hamiltonian. One introduces a probe gauge field
${\cal A}_{l_{x},l_{y};n_{x},n_{y}}$ (in addition to the physical
one), acting on links between the different sites. Taking as a concrete
but representative example the compatible-L case, Eqs. \eqref{eq:dissH}
and \eqref{eq:Hcomp}, the Hamiltonian is given by\begin{widetext}
\begin{align}
H & =-t_{{\rm diss}}\sum_{l_{x},l_{y},s=\pm1}b_{l_{x},l_{y}}^{\dagger}\left(e^{isl_{x}2\pi\alpha}e^{i{\cal A}_{l_{x},l_{y};l_{x},l_{y}+s}}a_{l_{x},l_{y}+s}+e^{i{\cal A}_{l_{x},l_{y};l_{x}+s,l_{y}}}a_{l_{x}+s,l_{y}}\right)\nonumber \\
 & -t_{{\rm comp}}\sum_{l_{x},l_{y}}a_{l_{x},l_{y}}^{\dagger}\left(e^{i{\cal A}_{l_{x},l_{y};l_{x},l_{y}+1}}e^{il_{x}2\pi\alpha}a_{l_{x},l_{y}+1}+e^{i{\cal A}_{l_{x},l_{y};l_{x}+1,l_{y}}}a_{l_{x}+1,l_{y}}\right)\nonumber \\
 & -\mu^{*}\sum_{l_{x},l_{y}}b_{l_{x},l_{y}}^{\dagger}a_{l_{x},l_{y}}+\mathrm{h.c.}+H^{{\rm bath}}
\end{align}
The particle currents $J^{i}$ are then given by taking the derivative
w.r.t. ${\cal A}$, and taking ${\cal A}\to0$, \begin {subequations}
\begin{align}
J_{l_{x},l_{y}}^{x} & =\frac{\delta H}{\delta{\cal A}_{l_{x},l_{y};l_{x}+1,l_{y}}}|_{{\cal A}\to0}=-it_{{\rm comp}}\left(a_{l_{x},l_{y}}^{\dagger}a_{l_{x}+1,l_{y}}-a_{l_{x}+1,l_{y}}^{\dagger}a_{l_{x},l_{y}}\right)-it_{{\rm diss}}\left(b_{l_{x},l_{y}}^{\dagger}a_{l_{x}+1,l_{y}}-a_{l_{x}+1,l_{y}}^{\dagger}b_{l_{x},l_{y}}\right),
\end{align}
\begin{align}
J_{l_{x},l_{y}}^{y} & =\frac{\delta H}{\delta{\cal A}_{l_{x},l_{y};l_{x},l_{y}+1}}|_{{\cal A}\to0}\nonumber \\
 & =-it_{{\rm comp}}\left(e^{il_{x}2\pi\alpha}a_{l_{x},l_{y}}^{\dagger}a_{l_{x},l_{y}+1}-e^{-il_{x}2\pi\alpha}a_{l_{x},l_{y}+1}^{\dagger}a_{l_{x},l_{y}}\right)-it_{{\rm diss}}\left(e^{il_{x}2\pi\alpha}b_{l_{x},l_{y}}^{\dagger}a_{l_{x},l_{y}+1}-e^{-il_{x}2\pi\alpha}a_{l_{x},l_{y}+1}^{\dagger}b_{l_{x},l_{y}}\right).
\end{align}
\end {subequations}\end{widetext}As can be easily seen, $J^{i}$
is the sum of an $a$-$a$ current, whose definition exactly coincides
with our definitions for the particle current in the main text, Eqs.
\eqref{eq:jxcomp},\eqref{eq:jycomp}, and of an $a$-$b$ contribution,
describing exchange of particles between the bath and the lattice
$a$-particles. The same procedure may be repeated for any choice
of $H^{{\rm sys}}$, arriving at a similar outcome. 

Let us now turn to evaluate the $a$-$b$ part, and show how it gives
rise to the dissipative current, Eq. \eqref{eq:ndotbrokenapart}.
For this purpose let us use the notation $H^{{\rm diss}}=\sum_{\alpha\beta}b_{\beta}^{\dagger}h_{\beta\alpha}^{{\rm ref}}a_{\alpha}+{\rm h.c.}$,
where $\alpha,\beta$ are generalized site indices. According to our
present derivation of the $J^{i}$ operators, the operator corresponding
to the total outgoing particle flux from a site $\alpha$ is
\begin{equation}
J_{\alpha}^{\mathcal{D}}=i\left[\sum_{\beta}b_{\beta}^{\dagger}\left(h_{\alpha\beta}^{{\rm ref}}\right)^{*}\right]a_{\alpha}-ia_{\alpha}^{\dagger}\left[\sum_{\beta}h_{\alpha\beta}^{{\rm ref}}b_{\beta}\right].\label{eq:JDoutExact}
\end{equation}
Eq. \eqref{eq:JDoutExact} is linear in the dissipative coupling,
so in order to be consistent with the derivation of the Lindblad equation
itself, which is second order in the system-bath coupling \citep{MG},
we should evaluate it to first-order in $h^{{\rm ref}}$ (remembering
that the zeroth-order term vanishes). Using the Kubo formula,
\begin{align*}
\left\langle J_{\alpha}^{\mathcal{D}}\left(t\right)\right\rangle  & =-i\int_{0}^{\infty}d\tau\left\langle \left[J_{\alpha}^{\mathcal{D}}\left(t\right),H^{{\rm diss}}\left(t-\tau\right)\right]\right\rangle ,
\end{align*}
where in inside the integral the averaging is performed in the $h^{{\rm ref}}\to0$
state. Expanding the commutator term inside of the integral, and using
the separability of the $\left\langle a^{\dagger}a\right\rangle $
and $\left\langle b^{\dagger}b\right\rangle $ correlations for $h^{\mathrm{ref}}=0$,
we find
\begin{align}
\left\langle J_{\alpha}^{\mathcal{D}}\left(t\right)\right\rangle  & =\sum_{\beta\alpha^{\prime}\beta^{\prime}}\int_{0}^{\infty}d\tau\left(h_{\alpha\beta}^{{\rm ref}}\right)^{*}h_{\alpha^{\prime}\beta^{\prime}}^{{\rm ref}}\times\nonumber \\
 & \Bigl[\left\langle b_{\beta}^{\dagger}\left(t\right)b_{\beta^{\prime}}\left(t-\tau\right)\right\rangle \left\langle a_{\alpha}\left(t\right)a_{\alpha^{\prime}}^{\dagger}\left(t-\tau\right)\right\rangle \nonumber \\
 & -\left\langle a_{\alpha^{\prime}}^{\dagger}\left(t-\tau\right)a_{\alpha}\left(t\right)\right\rangle \left\langle b_{\beta^{\prime}}\left(t-\tau\right)b_{\beta}^{\dagger}\left(t\right)\right\rangle \Bigr]\nonumber \\
 & +\sum_{\beta\alpha^{\prime}\beta^{\prime}}\int_{0}^{\infty}d\tau\left(h_{\alpha^{\prime}\beta^{\prime}}^{{\rm ref}}\right)^{*}h_{\alpha\beta}^{{\rm ref}}\times\nonumber \\
 & \Bigl[\left\langle b_{\beta^{\prime}}^{\dagger}\left(t-\tau\right)b_{\beta}\left(t\right)\right\rangle \left\langle a_{\alpha^{\prime}}\left(t\right)a_{\alpha}^{\dagger}\left(t-\tau\right)\right\rangle \nonumber \\
 & -\left\langle a_{\alpha}^{\dagger}\left(t\right)a_{\alpha^{\prime}}\left(t-\tau\right)\right\rangle \left\langle b_{\beta}\left(t\right)b_{\beta^{\prime}}^{\dagger}\left(t-\tau\right)\right\rangle \Bigr].
\end{align}
Now, we employ the fact that b-particles quickly evaporate after escaping
the 2D lattice, so that we may set $\left\langle b_{\alpha}^{\dagger}\left(t\right)b_{\beta}\left(t'\right)\right\rangle =0$.
In addition, by the Markov approximation used in deriving the Lindblad
equation, which implies that the bath dynamics is much faster than
the system dynamics, one may set $\left\langle b_{\alpha}\left(t\right)b_{\beta}^{\dagger}\left(t'\right)\right\rangle =\frac{2\pi}{\hbar}\nu_{0}\delta_{\alpha\beta}\delta\left(t-t'\right)$,
where $\nu_{0}$ is the density of states of the $b$-particles, as
introduced in Sec. \ref{dissScheme}. Plugging these identities in,
we obtain
\begin{align}
\left\langle J_{\alpha}^{\mathcal{D}}\right\rangle  & =-\frac{2\pi}{\hbar}\nu_{0}\sum_{\alpha^{\prime}\beta^{\prime}}\left[\left(h_{\alpha\beta^{\prime}}^{{\rm ref}}\right)^{*}h_{\alpha^{\prime}\beta^{\prime}}^{{\rm ref}}\left\langle a_{\alpha^{\prime}}^{\dagger}a_{\alpha}\right\rangle +\left(h_{\alpha^{\prime}\beta^{\prime}}^{{\rm ref}}\right)^{*}h_{\alpha\beta^{\prime}}^{{\rm ref}}\left\langle a_{\alpha}^{\dagger}a_{\alpha^{\prime}}\right\rangle \right],
\end{align}
and the time argument is suppressed, as all the operators are evaluated
at the same time. Performing the summation over the dummy index $\beta^{\prime}$,
and using the definition in Eq. \eqref{eq:Gamma1Gammout},
\begin{align}
\left\langle J_{\alpha}^{\mathcal{D}}\right\rangle  & =-\frac{1}{2}\sum_{\alpha^{\prime}}\left[\Gamma_{\alpha\alpha^{\prime}}^{\left(1\right)}\left\langle a_{\alpha^{\prime}}^{\dagger}a_{\alpha}\right\rangle +\Gamma_{\alpha^{\prime}\alpha}^{\left(1\right)}\left\langle a_{\alpha}^{\dagger}a_{\alpha^{\prime}}\right\rangle \right]\nonumber \\
 & =-\frac{1}{2}\left\{ \Gamma^{\left(1\right)},P\right\} _{\alpha,\alpha}.
\end{align}
This result reproduces exactly the second half of Eq. \eqref{eq:ndotbrokenapart}.
The first half, which originates in a constant refilling process captured
by $\Gamma^{\left(2\right)}$, is obtained in a similar way, by introducing
a refilling reservoir of $c$ atoms with infinitely positive chemical
potential. This would give us the remaining $\left\{ \Gamma^{\left(2\right)},1-P\right\} $
term. Thus, we have verified that the current operator we calculate
is indeed the physical current of particles within the lattice. (Despite
the seeming ambiguity in its derivation from the continuity equation,
adding a divergence-free field to $j^{x/y}$ would not correspond
to any measurable transport current of either particles.) 

\section {Definition of the conductance tensor}\label{appC}

\begin{figure*}
\begin{centering}
\includegraphics[bb=33.6059bp 0bp 919.361bp 322.504bp,scale=0.57]{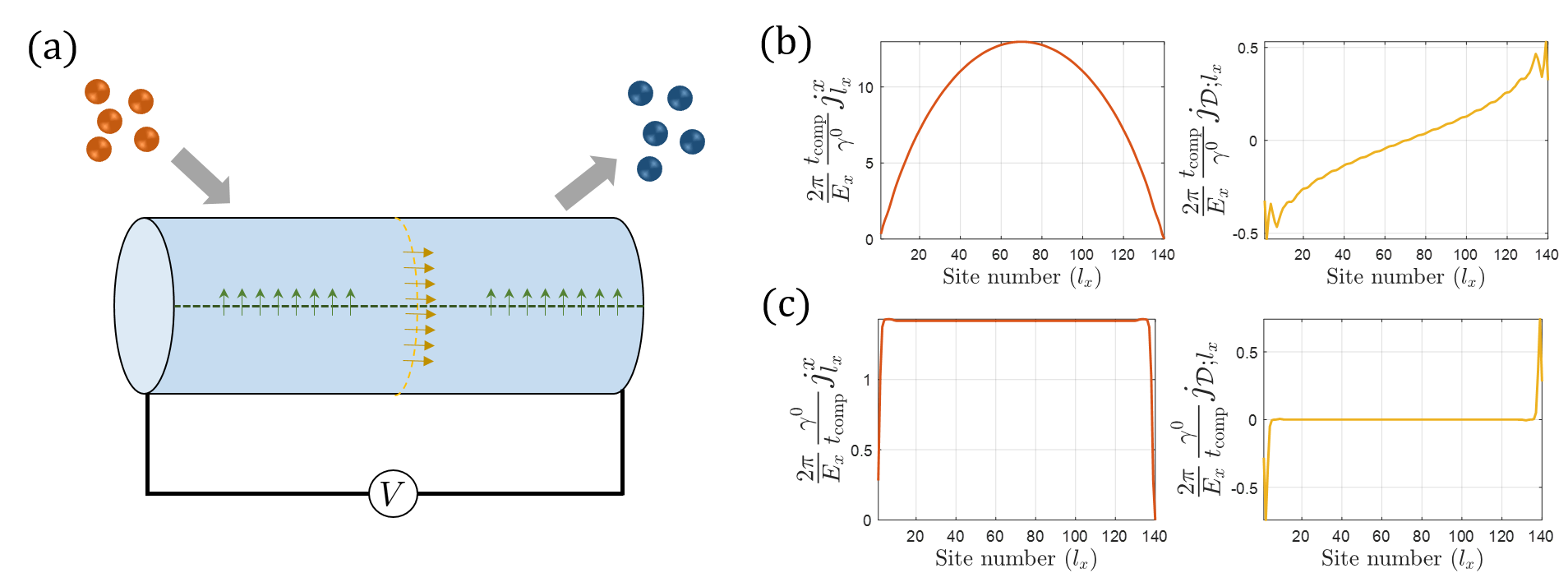}
\par\end{centering}
\caption{\label{fig:Conddefinitions} (a) Geometry of the cylindrical system
with applied potential differences between its ends. The cross section
corresponding to the calculation of the transverse response is marked
in green, and that of the longitudinal response appears in yellow.
The longitudinal current is generated by an incoming flux of bath
particles in one side and an outgoing one in the other side. (b) Profiles
of the longitudinal current profile (left, red) and dissipative current
(right, yellow) in the case of a dominant compatible Hamiltonian.
Here the dissipative current roughly follows the applied potential,
and the longitudinal current adjusts into a parabolic profile to obey
the conservation law, Eqs. \ref{eq:dndt20}--\ref{eq:dndt22}. In
the middle of the system, $j^{x}$ has zero slope and is a good measure
of the total flux of particles going between the two sides of the
bath. In this calculation we applied current along the $\hat{x}$
direction and used $t_{{\rm comp}}=1000\gamma^{0}$, $L_{x}=L_{y}=140$,
$\alpha=\frac{1}{7}$, and the optimal values for $\mu^{*}$ and $\gamma^{{\rm in}}$.
(c) The same as (b), in the dissipative regime, where $t_{{\rm comp}}=0.001\gamma^{0}$.
Here the flux of bath particles is localized near the edges, but $j_{\frac{1}{2}L_{x}}^{x}$
plays the same role. The results of calculating these current profiles
for the Hamiltonian-dominant and dissipative regime of the incompatible
case are qualitatively similar to (b) and (c), respectively (up to
additional weak oscillatory behavior in $l_{x}$ in the former).}
\end{figure*}

We elucidate in this appendix the definition and physical meaning
of the conductance we calculate in this work, as they appear in Eq.
\eqref{eq:condtensor}. We consider a cylindrical geometry, with periodic
boundary conditions applied perpendicular to the direction of applied
voltage (or more specifically electric field gradient in our case),
as demonstrated in Fig. \ref{fig:Conddefinitions}a.

For the transverse response, e.g., $G_{yx}$, we consider the total
current going around the circumference of the cylinder along a longitudinal
cut, averaged around the cylinder. As $G_{yx}$ is the ratio of total
current through this cross section to the voltage difference applied,
we divide the total current by the length times the applied field,
e.g., for $G_{yx}$ we divide by $L_{x}E_{x}$. 

The issue of longitudinal current is a bit more subtle, as the notion
of a particle current flowing between two leads does not exist in
the system we consider. Moreover, in the steady state the total particle
number in the system is conserved. We quantify our response by examining
the total flux of particles entering the system from one side and
leaving it in the other. As can be seen in Fig. \ref{fig:Conddefinitions}b--c,
the electric field gradient induces an incoming flux of bath particles
(negative $j_{{\cal D}}$) on the right side and an outgoing one (positive
$j_{{\cal D}}$) on the left. It is then sensible to define this response
by the longitudinal current exactly in the middle of the cylinder,
since it is equal in amplitude to the total dissipative current integrated
over one half of the system. 

\begin{figure}
\begin{centering}
\includegraphics[bb=0bp 0bp 398bp 186bp,scale=0.62]{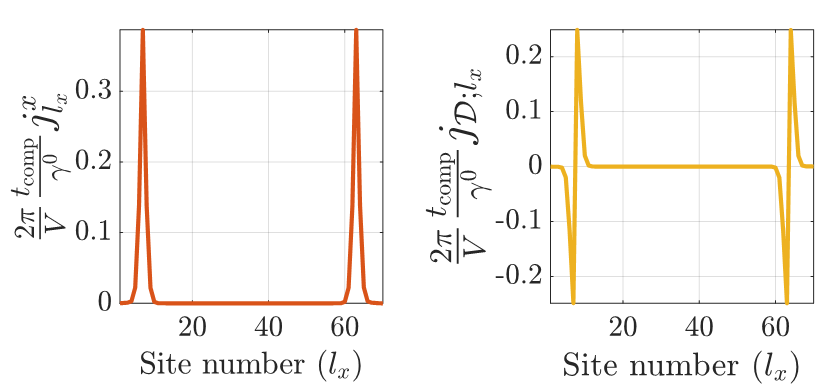}
\par\end{centering}
\caption{\label{fig:leadsprofiles} Example of longitudinal current profiles
with the perturbation $\delta H_{{\rm leads}}$. The longitudinal
current (left, red) and the dissipative current (right, yellow) are
calculated with the system Hamiltonian $H_{{\rm comp}}^{{\rm sys}}$,
in the dissipative regime, $t_{{\rm comp}}=0.001\gamma^{0}$. We used
$L_{x}=L_{y}=70$, $\alpha=\frac{1}{7}$, $L_{{\rm lead}}=7$, and
the optimal values for $\mu^{*}$ and $\gamma^{{\rm in}}$ in this
calculation. Notice the response is extremely localized near the leads.
We have verified that the same qualitative localization occurs for
all regimes and $H^{{\rm sys}}$ choices explored in this work.}
\end{figure}

Lastly, let us comment on a different method of applying a voltage
difference between the two edges. Namely, imposing ``voltage leads''
on the system, by imposing a finite positive chemical potential on
a narrow region near one edge, and a negative on near the other edge.
The perturbation to the Hamiltonian can be written as $\delta H_{{\rm leads}}=\frac{V}{2}\sum_{l_{y}}\delta H_{l_{y}}$,
with $V$ the applied voltage,
\begin{equation}
\delta H_{l_{y}}=\left(\sum_{l_{x}=1}^{L_{{\rm lead}}}-\sum_{l_{x}=L_{x}-L_{{\rm lead}}+1}^{L_{x}}\right)a_{l_{x},l_{y}}^{\dagger}a_{l_{x},l_{y}},
\end{equation}
and $L_{{\rm lead}}$ is the size of the leads. As illustrated in
Fig. \ref{fig:leadsprofiles}, the longitudinal current in this case
flows only at the vicinity of the perturbation. Moreover, the balanced
flux of particles in and out of the bath (captured by the dissipative
current) is localized as well, as bath particles enter and leave the
system in a very narrow region. In this case, there is no sense of
particle flow between the two sides of the system or bath, and it
appears $\delta H_{{\rm leads}}$ is not a good choice for studying
the longitudinal current. We note that for the transverse conductance
however, one recovers qualitatively the results of using $\delta H_{\boldsymbol{E}}$,
including the conductance quantization at the appropriate parameter
regime.

\end {appendix}

\bibliographystyle{../apsrev4-1}
\bibliography{../../references2}

\end{document}